%% file: main.tex
\documentclass[journal]{IEEEtran}

\usepackage[T1]{fontenc}
\usepackage{graphicx}
\graphicspath{ {./Figs/} }
\usepackage{physics}
\usepackage[ruled,linesnumbered]{algorithm2e}
\RestyleAlgo{ruled}
\SetKwComment{Comment}{/* }{ */}
\usepackage{multirow}
\usepackage{pifont}
\usepackage{enumitem}

\ifCLASSINFOpdf
\else
\fi

\usepackage{amsmath}
\usepackage[cmintegrals]{newtxmath}
\usepackage{bm}

\usepackage{soul}

\begin{document}

\title{Qu-Trefoil: Large-scale Quantum Circuit Simulator Working on FPGA with SATA Storages}

\author{Kaijie~Wei,~\IEEEmembership{Member,~IEEE,}
        Hideharu~Amano,~\IEEEmembership{Life Member,~IEEE,}
        Ryohei~Niwase,~\IEEEmembership{Member,~IEEE,}
        Yoshiki~Yamaguchi,~\IEEEmembership{Member,~IEEE,}
        and~Takefumi~Miyoshi
\thanks{K.~Wei is with Center for Sustainable Quantum AI, Keio University, 2-15-45 Mita, Minato City, Tokyo, Japan 108-0073.
\\E-mail: wei@am.ics.keio.ac.jp 
}
\thanks{H.~Amano is with Systems Design Lab, The University of Tokyo, 2-11-16 Yayoi, Bunkyo-ku, Tokyo, Japan 113-0032.
\\E-mail: hunga@dlab.t.u-tokyo.ac.jp 
}
\thanks{R.~Niwase and Y.~Yamaguchi are with the Graduate School of Systems and Information Engineering, University of Tsukuba, 1-1-1 Ten-ou-dai Tsukuba Ibaraki, Japan 305-8573.
\\E-mail: \{niwase,~yamaguchi.yoshiki.ge\}@lila.cs.tsukuba.ac.jp
}
\thanks{T.~Miyoshi is with WasaLabo, LLC., Blanche K-2, 6-5-20 Minaminaruse Machida, Tokyo, Japan, 194-0045.
\\E-mail: miyo@wasamon.net}
}

\markboth{IEEE TRANSACTIONS ON COMPUTERS,~Vol.~14, No.~8, August~2024}%
{K.~Wei \MakeLowercase{\textit{et al.}}: Qu-Trefoil: Large-scale Quantum Circuit Simulator Working on FPGA with SATA Storages}

\maketitle

\input{Contents/0_abst}

\IEEEpeerreviewmaketitle

\input{Contents/1_intro}
\input{Contents/2_back}
\input{Contents/3_trefoil}
\input{Contents/4_impl}
\input{Contents/5_eval}
\input{Contents/6_con}
\input{Contents/ack}

\ifCLASSOPTIONcaptionsoff
  \newpage
\fi

\bibliographystyle{IEEEtran}
\bibliography{main}

\input{Contents/bio}
\end{document}

%% file: Contents/0_abst.tex
\begin{abstract}
 
Quantum circuits are fundamental components of quantum computing, and state-vector-based quantum circuit simulation is a widely used technique for tracking qubit behavior throughout circuit evolution. However, simulating a circuit with $n$ qubits requires $2^{n+4}$ bytes of memory, making simulations of more than 40 qubits feasible only on supercomputers. To address this limitation, we propose the Qu-Trefoil, a system designed for large-scale quantum circuit simulations on an FPGA-based platform called Trefoil. Trefoil is a multi-FPGA system connected to eight storage subsystems, each equipped with 32 SATA disks. Qu-Trefoil integrates a suite of HLS-based universal quantum gates, including Clifford gates (Hadamard (H), Pauli-Z (Z), Phase (S), Controlled-NOT (CNOT)), the T gate, and unitary matrix computation, along with HDL-designed modules for system-wide integration. Our extensive evaluation demonstrates the system’s robustness and flexibility, covering quantum gate performance, chunk size, disk extensibility, and efficiency across different SATA generations. We successfully simulated quantum circuits with over 43 qubits, which required more than 128 TB of memory, in approximately 3.72 to 13.06 hours on a single storage subsystem equipped with one FPGA. This achievement represents a significant milestone in the advancement of quantum computing simulations. Furthermore, thanks to its unique architecture, Qu-Trefoil is more accessible, flexible, and cost-efficient than other existing simulators for large-scale quantum circuit simulations, making it a viable option for researchers with limited access to supercomputers.

\end{abstract}


%% file: Contents/1_intro.tex
\section{Introduction}
\label{sec: intro}

\IEEEPARstart{Q}{uantum} computing represents a revolutionary paradigm that leverages the principles of quantum mechanics to solve problems far beyond the capabilities of classical computers. Unlike classical computing, where bits represent 0 or 1, quantum bits (qubits) can exist in superposition, representing both 0 and 1 simultaneously. This characteristic allows quantum computers to perform massively parallel computations. Research on quantum algorithms, including Shor's algorithm \cite{shor}, Grover's algorithm \cite{grover}, Quantum Phase Estimation (QPE) \cite{QPE}, and others, has garnered worldwide attention due to their exponential speedup, far surpassing classical computing. According to the latest report from IBM, it has unveiled an 1121-qubit quantum processor, Condor, and plans to develop a 10,000-qubit quantum computer by 2029 \cite{ibm_latest}. However, handling such large-scale qubits remains highly challenging, even for IBM, which stems from issues like mapping problems onto the machine's topology and controlling all qubits simultaneously due to noise and coherence time limitations. On the other hand, quantum circuit simulation is critical for developing and testing quantum algorithms because of several limitations in the current state of quantum computers, including noise, scalability, limited qubit connectivity, and others \cite{quantum_challenegs}.

There are three primary types of quantum simulators: density matrix \cite{DM_intro}, tensor network \cite{TN_intro}, and state vector (SV) \cite{SV_app_1}. Density matrix simulators can simulate mixed states and account for noise and decoherence. On the other hand, tensor network simulators leverage the entanglement structure of quantum states to reduce computational complexity, making them a favorable option for handling larger circuits. Lastly, SV-based quantum circuit simulators (QCSs) provide researchers with practical tools for debugging and comprehending quantum circuits, as they allow for an explicit representation of quantum states, a popular method among researchers studying their algorithms.

Simulating large-scale quantum circuits using SV presents significant challenges, particularly regarding resource requirements and simulation speed. Simulators become impractical as the number of qubits increases due to the exponential resource requirement of $2^{n+4}$ bytes for $n$ qubits, where each state vector is stored using double-precision floating-point complex numbers. The representations of double-precision floating-point are especially critical for applications requiring long-term coherence or involving highly sensitive calculations, such as quantum error correction (QEC), phase estimation, and fault-tolerant quantum computing, where even minor numerical errors can accumulate and lead to incorrect results compared to single-precision or fixed-point representations.

Moreover, simulating quantum circuits at the scale of state-of-the-art quantum computers is infeasible on classical simulators due to the exponential resource demands. The complexity of such systems far exceeds what even the most advanced classical supercomputers can handle, highlighting the growing gap between classical hardware simulation capabilities and the actual performance of modern quantum computers.

Despite these limitations, recent advancements in QCS on supercomputers have made significant progress, targeting qubit scales around 40 to 50, which are sufficient to capture the core behavior and performance characteristics of algorithms like Grover's algorithm or the quantum approximate optimization algorithm rather than requiring larger scales of over 100 qubits. For example, the Fugaku supercomputer successfully simulated 48 qubits \cite{fugaku}, and ARCHER2, which achieved 44 qubits using QuEST \cite{archer}. In 2023, Nvidia Crop simulated 40 qubits on the Selene supercomputer using Qiskit Aer with 256 NVIDIA DGX A100s \cite{cuQuantum}. However, the simulated qubit scale for a Graphics Processing Units (GPU) server like NVIDIA DGX H100 is limited to 33 \cite{nvidia}. Therefore, while supercomputers and high-performance computing (HPC) clusters equipped with multiple GPUs dominate the simulation of quantum circuits with more than 35 qubits, accessibility and investment remain significant challenges for researchers, making large-scale simulations difficult to achieve for many academic and industrial institutions.

Field-programmable gate arrays (FPGAs) are known for their flexibility and energy efficiency, but their limited memory capacity and bandwidth have hindered their further application in quantum circuit simulation. Some FPGAs provide powerful High Bandwidth Memory (HBM)/HBM2 \cite{HBM_1, HBM_2}. However, the memory volumes are about 4GB$\sim$32GB, encountering memory issues for qubits over 35. Additionally, the quantum simulation itself is not a computationally intensive job, and most gate computations involve simple data interchange or coefficient multiplication, which DSP modules are enough on mid-range FPGA families \cite{xilinx, IntelAgilex}. To address the challenge of the platform, we propose a system working on a platform directly connecting FPGA with Serial Advanced Technology Attachment (SATA) disks named Tightly-coupled REconfigurable Fork Inter Link (Trefoil) \cite{MCSOC}.

\begin{figure}[ht]
    \centering
    \includegraphics[width=1.0\linewidth]{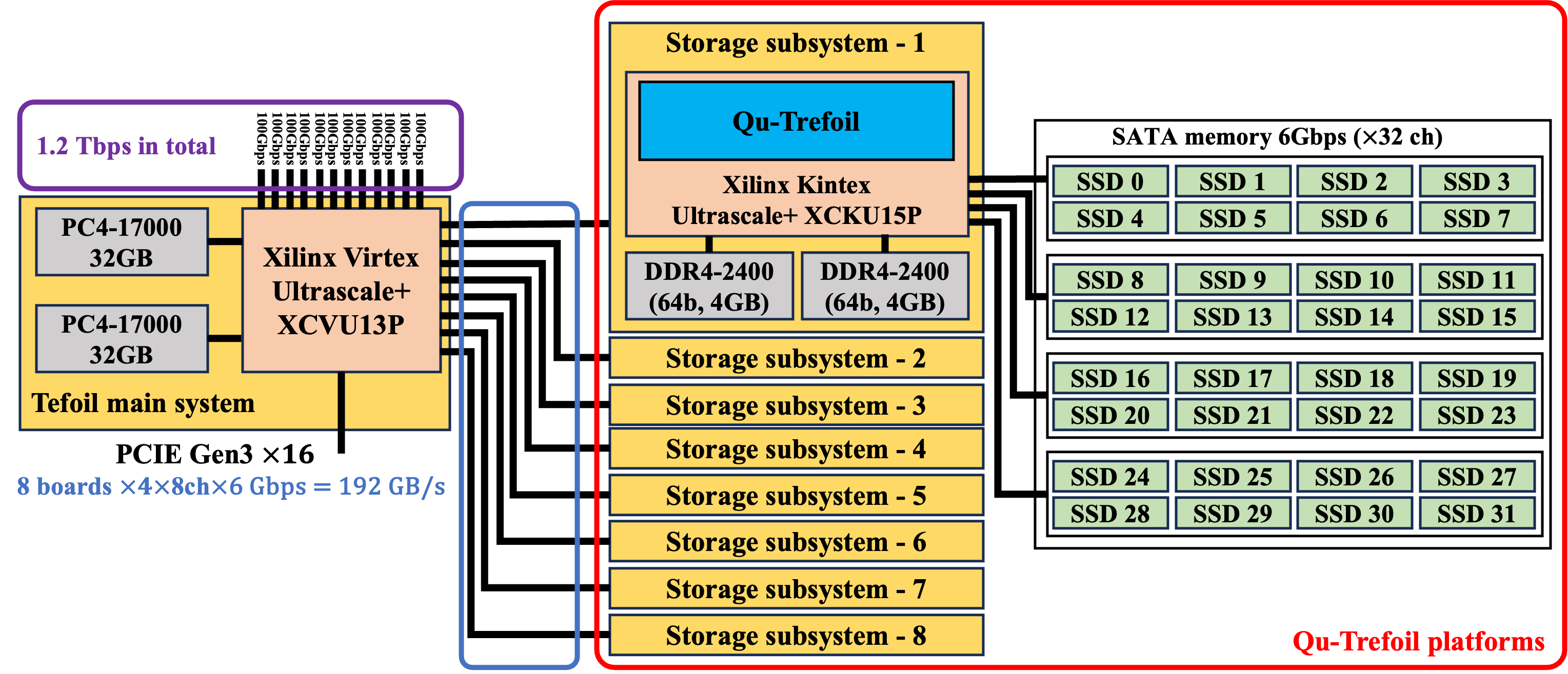}
    \caption{Trefoil's main system connecting to eight storage subsystems, offering 1.2 Tbps network access, 192GB/s storage throughput, and a total capacity of 2 PBytes using 8 TB SATA disks; Each storage subsystem equipped with an embedded Qu-Trefoil system}
    \label{fig: overall_sys}
\end{figure}

Trefoil is a low-cost, energy-efficient FPGA-based storage system designed for scalable storage and access performance through parallel processing with multiple FPGAs, as illustrated in Fig.~\ref{fig: overall_sys}. The main system, Virtex Ultrascale+XCVU13P, features abundant computational resources connected to the storage subsystems. Our proposed system, Qu-Trefoil, operates on a storage subsystem connected to 32 SATA disks, each with 8 TB of storage.

On the other hand, to implement the Qu-Trefoil, we utilize \cite{Qulacs}, a powerful Python/C++ library designed for simulating large, noisy, or parametric quantum circuits. Qulacs supports Central Processing Unit (CPU) and Graphics Processing Unit (GPU) platforms, with optimization features such as SIMD and OpenMP parallelization.

We summarize the existing challenges in integrating the above two ingredients three-fold. 
\begin{itemize}
\item The design of Qulacs does not consider the characteristics of FPGA, let alone the customized storage subsystem. 
\item The limited on-chip resources pose a challenge: As more SATA disks are involved, the resource utilization increases proportionally. Therefore, it is crucial to weigh the trade-off between on-chip resources and simulator performance. 
\item The slow data accessing speed of SATA disks comes to be the most significant bottleneck, especially when juxtaposed against the high-performance memory systems of supercomputers.
\end{itemize}

To address these issues, we propose a reliable system, Qu-Trefoil, that implements a set of universal quantum gates, allowing the construction of any arbitrary quantum circuit. The primary technical contributions of the proposed system are summarized as follows,
\begin{itemize}
\item We propose an FPGA-based design to facilitate the simulation scale over 40 qubits, which is the first practical attempt to use FPGAs for SV-based quantum circuit simulation on such a large scale, typically working on supercomputers-scale platforms.
\item We leverage the unique characteristics of SATA disks, employing burst-mode data transfer within the system to enhance data throughput and improve overall simulation performance.
\item We optimize the quantum gate implementations by utilizing high-level synthesis (HLS) techniques, carefully balancing the performance and resource utilization of the target FPGA platform.
\item We provide practical implementation details and a comprehensive simulator evaluation, including real-world performance metrics such as runtime and resource utilization.
\end{itemize}

Besides, the Qu-Trefoil system control process unfolds as follows,
\begin{enumerate}
\item The user loads a bitstream file onto a Trefoil storage subsystem via the JTAG port, configuring the FPGA based on the quantum gate and target qubit within the quantum circuit;
\item The user inputs parameters, such as qubit scale and target qubit, using a Python-based GUI on a Raspberry Pi3 connected to a target platform;
\item The system then performs the quantum gate computation, saving the results across 32 SATA disks;
\item Finally, the user retrieves the results from the SATA disks and outputs them to the Raspberry Pi3 for data visualization.
\end{enumerate}

The structure of this article is as follows: Section~\ref{sec: back} covers the necessary background of quantum computers and the state-of-the-art quantum simulator. Section~\ref{sec: trefoil} specifies the details of the target platform. We elaborate on the design of Qu-Trefoil in Section~\ref{sec: impl} and evaluate the performance by exploiting the target platform through various experiments described in Section~\ref{sec: eval}. Finally, we conclude and discuss our work in Section \ref{sec: con}.

%% file: Contents/2_back.tex
\section{Background and Motivation}
\label{sec: back}

This section first introduces the fundamental concepts and hurdles associated with quantum computers. Then, we provide a comprehensive overview of the SV-based QCS, covering the cutting-edge designs and the motivation of this research. Finally, we describe one of the representative ones in detail, the base design of our system, Qulacs.

\subsection{Quantum Computing}
\label{subsec: QC}
Quantum computers operate based on the principles of quantum mechanism, which can revolutionize various fields, including drug discovery, materials science, cryptography, and more, by solving complex problems that currently challenge classical computers \cite{Q_app}. Unlike classical computers, which use bits as the fundamental unit of information (either 0 or 1), quantum computers use qubits in a superposition of 0 and 1 states, allowing them to process vast amounts of information in parallel virtually. However, several challenges still limit their practical and widespread adoption. Fidelity is a persistent issue, as quantum computers are susceptible to noise, decoherence, and errors that can disrupt the fragile quantum states of qubits. Scalability is another challenge because maintaining long-range connectivity and minimizing crosstalk between qubits becomes increasingly tricky with an increasing number of qubits. Finally, the challenge of hardware complexity and cost is ever-present since building and operating quantum computers require advanced techniques and sophisticated cryogenic systems to maintain ultra-low temperatures. Despite significant progress made in recent years, realizing the full potential of quantum computing in solving real-world problems will take time and effort.

Therefore, QCSs have been crucial tools for developing, testing, and understanding quantum algorithms and circuits, allowing researchers to simulate quantum systems' behaviours without accessing the physical quantum computers.

\subsection{Quantum Circuit Simulation}
In recent years, the development of QCSs has made significant progress. \textbf{SV}, \textbf{density matrix}, and \textbf{tensor network} have been widely adopted for different purposes and applications \cite{SV_app_1, DM_app_1, TN_app_1}. \textbf{SV} have taken notable positions in understanding quantum circuits by explicitly representing quantum states. \textbf{Density matrix}, on the other hand, allows the simulation of mixed states and incorporates the effects of noises and decoherence despite the intensive computational complexity. Lastly, \textbf{tensor network} is a practical tool for compactly representing the entanglement structure while losing some detailed information in the circuits.

We chose SV as the target method in our system because it is the most accurate method for describing the details in a quantum circuit with "ideal" noiseless qubits, making it superior to the other two methods in \textbf{quantum circuit debugging}. In representing the superposition of these states, 
\begin{equation*}
\ket{\psi} = \alpha \ket{0} + \beta \ket{1}
\end{equation*}
denotes the state of a single qubit, where $\ket{0}$ and $\ket{1}$ are ket notation to represent column vectors $(0, 1)$ and $(1, 0)$, respectively. $|\alpha|^2$ and $|\beta|^2$ represent the probabilities of the above two kets, obeying the rule of $|\alpha|^2+|\beta|^2=1$. 
For a quantum circuit with $n$ qubits, the state can be expressed as: 
\begin{equation*}
\ket{\psi} = \alpha_{0\cdots 00} \ket{0\cdots 00} + \alpha_{0\cdots 01} \ket{0\cdots 01} + \cdots +\alpha_{1\cdots 11} \ket{1\cdots 11}
\end{equation*},
which represents a vector with $2^n$ complex numbers, where the amplitudes are stored as double-precision floating-point values. Since each complex number requires 16 bytes (8 bytes each for the real and imaginary parts), storing this state requires $2^n \times 16 = 2^{n+4}$ bytes of memory, as introduced in Section \ref{sec: intro}. Consequently, the exponential memory requirement becomes a significant challenge as the number of qubits increases. On the other hand, in an SV-based QCS, the fundamental operation is the quantum gate, and a sequence of such gates forms a quantum circuit. Applying a quantum gate to a quantum state during simulation is equivalent to performing a matrix-vector multiplication, which reflects how quantum operations alter the state at each step.

\subsection{State-vector-based Quantum Circuit Simulator}
\label{subsec: QCS}
Concerning the simulator development, as one of the most representative SV-based QCSs, Intel Quantum Simulator (IQS), an open-source environment for HPC and cloud computing infrastructures,  presents impressive benchmarks for large-scale simulations of up to 42 qubits employing 4096 processes, with 2048 nodes of the SuperMUC-NG system \cite{IQS}. QuEST \cite{quest}, embodied as a C library, simulates circuits of up to 38 qubits distributed over 2048 compute nodes, each with up to 24 cores using the ARCUS Phase-B and ARCHER supercomputers. The NVIDIA cuQuantum SDK \cite{cuQuantum} provides a GPU-accelerated implementation, realizing a 40-qubit simulation with 128 GPUs on the H100 cluster \cite{nvidia}. Finally, Qulacs \cite{Qulacs}, the base algorithm of Qu-Trefoil, supports three modes, including OpenMP mode for servers, SIMD mode for SIMD extension, and GPU mode for GPU accelerators. We will discuss Qulacs further in Section \ref{subsec: qulacs} correlating to our design. Despite the significant progress in simulator designs, the costs and the accessibility of target platforms to researchers impede the development of quantum computing. Additionally, the simulation of quantum circuits on a large qubit scale requires handling enormous memory capacity and data communication.

On the other hand, from the aspect of the platform, GPUs and CPUs are commonly adopted, as discussed previously. Although there are many reports on implementing annealers or Ising machines on FPGAs \cite{annealing_1, annealing_2, Ising}, these researches have yet to delve into the topics of quantum gate simulation. On the other hand, \cite{FPGA_Qcomp_1} and \cite{FPGA_Qcomp_2} have focused on the computational aspects of quantum gates, using HDL description, but have not discussed the memory system for storing SVs, which is a crucial issue for FPGA implementation. Furthermore, in our earlier reports \cite{FPT_1} and \cite{FPT_2}, we presented preliminary results on small-scale qubits without dedicated optimizations, demonstrating the feasibility of the target system.

\subsection{Quantum Gates Implementations on Qulacs}
\label{subsec: qulacs}
Qulacs is a highly regarded QCS known for its speed \cite{Qulacs}. It is written in C and C++ and has a Python interface for ease of use. Many researchers have used Qulacs to accelerate their quantum computing research, including Fujitsu, who has developed a distributed SV-based QCS based on Qulacs called mpiQulacs. According to their report, they have evaluated up to 36 qubits on the Todoroki cluster, consisting of 64 nodes based on A64FX \cite{mpiQulacs} using mpiQulacs. The design emphasizes the inter-node exchange using MPI. 

As a vital component of the Qu-Trefoil design, we highlight the implementation of a set of universal quantum gates, which can form any other quantum gates, and a two-qubit operation called \textbf{double\_qubits\_dense\_matrix ($Matrix$)} in Qulacs, which involves unitary matrix multiplication and requires significant computational resources. The set of universal quantum gates includes Clifford gates such as \textbf{Hadmard ($H$)}, \textbf{Pauli-Z ($Z$)}, \textbf{Phase ($S$)}, and \textbf{Controlled-NOT ($CNOT$)}, along with the \textbf{T} and the \textbf{Unitary Matrix ($Matrix$)}, as summarized in TABLE~\ref{tab: gates}.

\begin{table}[htb]
\footnotesize
\vspace*{-0.6em}
\caption{Implemented quantum gates}\label{tab: gates}
\vspace*{-2.2em}
\begin{equation*}
\begin{array}{||c||c|c||}
\hline
\textrm{\textbf{Gate}}& 
\textrm{\textbf{Effect}}& 
\textrm{\textbf{Matrix}} \\
\hline\hline
\begin{array}{c}
{\quad}\\
\mbox{\boldmath $H$}\\
{\quad}\\
\end{array}
&
\begin{array}{cc}
\textrm{Creating superposition}\\
\textrm{(halfway between the $\ket{0}$ and $\ket{1}$)}
\end{array}
&
\dfrac{1}{\sqrt{2}} 
\begin{pmatrix}
   1 & 1 \\
   1 & -1 \\
\end{pmatrix}\\
\hline
\begin{array}{c}
{\quad}\\
\mbox{\boldmath $Z$}\\
{\quad}\\
\end{array}
&
\begin{array}{cc}
\textrm{A phase flip to a qubit}
\end{array}
&
\begin{pmatrix}
   1 & 0 \\
   0 & -1\\
\end{pmatrix}\\
\hline
\begin{array}{c}
{\quad}\\
\mbox{\boldmath $S$}\\
{\quad}\\
\end{array}
&
\begin{array}{cc}
\textrm{A phase factor of $i$ to $\ket{1}$}
\end{array}
&
\begin{pmatrix}
   1 & 0 \\
   0 & i\\
\end{pmatrix}\\
\hline
\begin{array}{c}
{\quad}\\
\mbox{\boldmath $CNOT$}\\
{\quad}\\
\end{array}
& 
\begin{array}{cc}
\textrm{Fliping the Target qubit} \\
\textrm{(Control qubit == $\ket{1}$)} \\
\end{array}
&
\begin{pmatrix}
1 & 0 & 0 & 0 \\
0 & 1 & 0 & 0 \\
0 & 0 & 0 & 1 \\
0 & 0 & 1 & 0 \\
\end{pmatrix}
\\
\hline
\begin{array}{c}
{\quad}\\
\mbox{\boldmath $T$}\\
{\quad}\\
\end{array}
&
\begin{array}{cc}
\textrm{A phase factor of $\frac{1+i}{\sqrt{2}}$ to $\ket{1}$}
\end{array}
&
\begin{pmatrix}
   1 & 0 \\
   0 & e^{i\pi/4}\\
\end{pmatrix}\\
\hline
\begin{array}{c}
{\quad}\\
\mbox{\boldmath $Matrix$}\\
{\quad}\\
\end{array}
&
\begin{array}{c}
\textrm{$4\times4$ unitary matrix on 2 qubits}
\end{array} &
\\
\hline
\end{array}
\end{equation*}
\vspace*{-0.7em}
\end{table}

%% file: Contents/3_trefoil.tex
\section{Hardware Platform: TREFOIL}
\label{sec: trefoil}

Trefoil \cite{trefoil}, a versatile FPGA platform manufactured by Tokyo Electron Device, is a highly adaptable FPGA solution that can operate autonomously in Multi-access Edge Computing (MEC) and other edge computing environments without relying on the host workstation. As discussed in Section \ref{sec: intro}, the system consists of a main system and eight storage subsystems. This paper emphasizes the QCS implementations on the Trefoil storage subsystem, which we will provide more details on the target platform.

\begin{figure}[ht]
    \centering
    \includegraphics[width=0.9\linewidth]{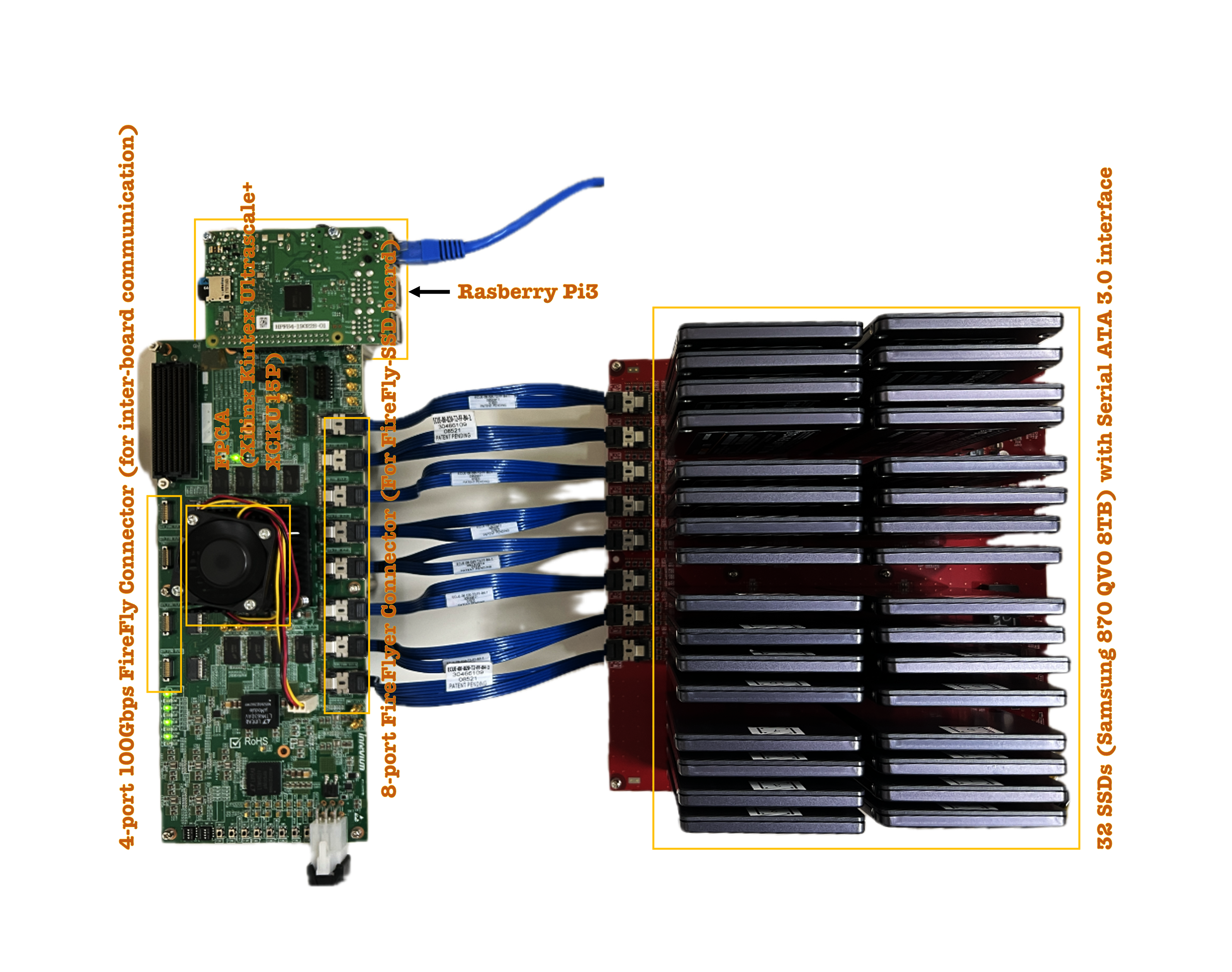}
    \vspace{-5mm}
    \caption{Trefoil's storage subsystem phototype}
    \label{fig: trefoil_storage}
\end{figure}

As the primary platform of Qu-Trefoil's design, the Trefoil storage subsystem can be adopted solely in edge computing systems. In Fig.~\ref{fig: trefoil_storage}, we present the prototype of this storage subsystem consisting of a subsystem FPGA platform and a FireFly-SSD board.

When discussing the storage subsystem's FPGA platform, which utilizes the Xilinx Kintex Ultrascale+ XCKU15P (with 1,143K system logic cells and 70.6Mb of memory), the 8-port FireFly connector facilitates data communication between the FireFly-SSD board and the FPGA, while the 4-port 100Gbps FireFly connector enables communication between the main system and the storage subsystem. The FireFly-SSD board is the primary interface between the SATA disks and the FPGA platform within the storage subsystem. Additionally, as shown in the top-left section of Fig.~\ref{fig: trefoil_storage}, a Raspberry Pi 3, functioning as the processing system (PS) of the platform, is connected for system control and data visualization, as introduced in Section~\ref{sec: intro}.

Regarding the storage device, serving as the storage capacity for the simulator, SATA Solid state drives (SSDs) perform significantly faster than traditional hard disk drives (HDDs). The Non-Volatile Memory Express (NVMe) standard \cite{disk_comparsion} outperforms the SATA standard in transfer speeds, but it requires four serial I/Os per NVMe SSD compared to only one for a SATA SSD. This trade-off allows the SATA standard to accommodate a more extensive system capacity from the limited FPGA I/Os standpoint. Opting for the SATA standard also leverages the simplified circuitry achievable with FPGA, as it eliminates the need for complex PCIe control required by NVMe SSDs, thereby enhancing stability and other associated benefits.

Furthermore, SATA-based SSDs present a more economical option. Consequently, following a thorough analysis, this paper selected a storage system utilizing SATA SSDs. We adopt Samsung 870 QVO \textbf{8TB} \cite{samsung_ssd} for storing SVs. In addition, although SATA disks read/write data at the unit of sector (512 bytes), we can considerably enhance performance by leveraging burst mode and the generations supported by the SATA interface \cite{samsung_ssd}, which we will evaluate in Section~\ref{subsec: eval_burst} and ~\ref{subsec: eval_gen}.

Speaking of the storage subsystem, 32 SATA disks provide \textbf{256TB} ($32 \times 8\textrm{TB}$) memory capacity for SV storage of up to \textbf{43 qubits}, achieving storage access speed of up to 24GBps ($4 \times 8\textrm{ch} \times 6\textrm{Gbps}$). As illustrated in Fig.~\ref{fig: overall_sys}, the main system can connect with eight storage subsystems. Thus, speaking of a full-geared Trefoil system, \textbf{2PB} ($8 \textrm{ boards}\times 256\textrm{TB}$) can be available for \textbf{46-qubit} simulation at the storage access speed of 192 GBps ($8 \textrm{ boards} \times 4 \times 8\textrm{ch} \times 6\textrm{Gbps}$).

%% file: Contents/4_impl.tex
\section{Hardware Implementation for Qu-Trefoil}
\label{sec: impl}

This section outlines the implementation process for Qu-Trefoil on FPGA, which involves three distinct phases. Firstly, we present the IP design, which takes charge of the data communication between SATA disks and FPGA using LiteX. Then, we discuss the intellectual property (IP) designs of the target set of universal quantum gates, as introduced in Section \ref{subsec: qulacs}. We use high-level synthesis (HLS) instead of hardware description languages (HDLs) like Verilog or VHDL to ensure efficient prototyping and maintenance of the simulator. Finally, we describe the overall system design of Qu-Trefoil, considering SATA disk access patterns in different scenarios of a target qubit.

\subsection{SATA disks $\Leftrightarrow$ FPGA Data Communication}
\label{subsec: SATA_FPGA}
Since Xilinx does not provide open-source IP for disk controllers, we can hardly realize the data communication between SATA disks and the FPGA. To solve this backward, we use an open-source tool, LiteX \cite{litex}, to provide the disk controller for Qu-Trefoil. 

LiteX is an open-source FPGA system integration tool developed and maintained by Enjoy-Digital on GitHub \cite{litex_github}. The Python-based tool can generate HDL, constraint files, and scripts for building, simulating, and debugging, providing the target script that describes the system configuration and platform script defining the FPGA vendor/parts/boards. More specifically, the target script and circuit description of the IP core provided by LiteX are written in Migen, an internal DSL (Domain-specific Language) of Python, offering a higher abstraction level than HDL-based system design \cite{migen_github}.

We use LiteSATA, an open-source SATA controller accompanying LiteX, as the IP core for controlling the data communication between FPGA and SATA disks. To allow multi-disk access, we have integrated a flexible framework that enables changing the number of SATA IP cores implemented on the FPGA and the number of data striping connections by specifying the number of storage devices as the build option from the command line. LiteSATA (tag: 2022.12) supports the high-speed transceivers (GTHE3\_CHANNEL and GTHE4\_CHANNEL) of Xilinx Kintex/Vertex Ultrascale(+), and it operates at SATA II (3.0 Gbps) and SATA III (6.0 Gbps) with 48-bit Logical Block Address (LBA) sector addressing. It consists of "Frontend," which includes the configurable crossbar,  stripping module, etc., "Core" (LiteSATACore), combining the physical layer with the data, transport, and command layers, and "Phy" (LiteSATAPHY) for handling the physical layer and SATA disks.

\begin{figure}[htp]
    \centering
    \includegraphics[width=0.9\linewidth]{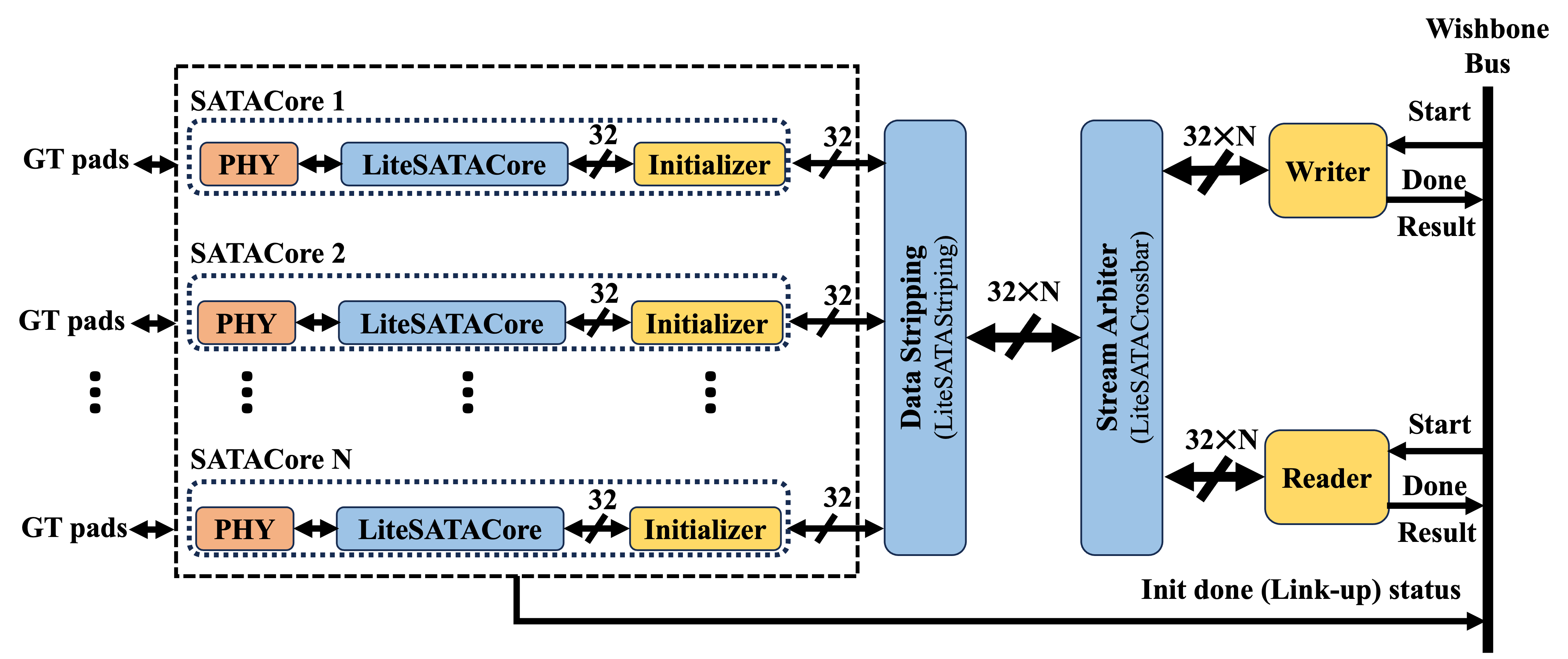}
    \caption{Sample design for LiteSATA working on multiple LiteSATACores}
    \label{fig: satacore}
\end{figure}
As illustrated in Fig.~\ref{fig: satacore}, our framework allows for simultaneous access to multiple LiteSATACore modules by designating the SATA ports to use in the system, which LiteX does not officially support. Moreover, we revised the initialization part and automated the process for each module, enabling efficient link-up detection and hot-swap functionality for multiple LiteSATACores. 

\begin{figure}[htp]
    \centering
    \includegraphics[width=0.8\linewidth]{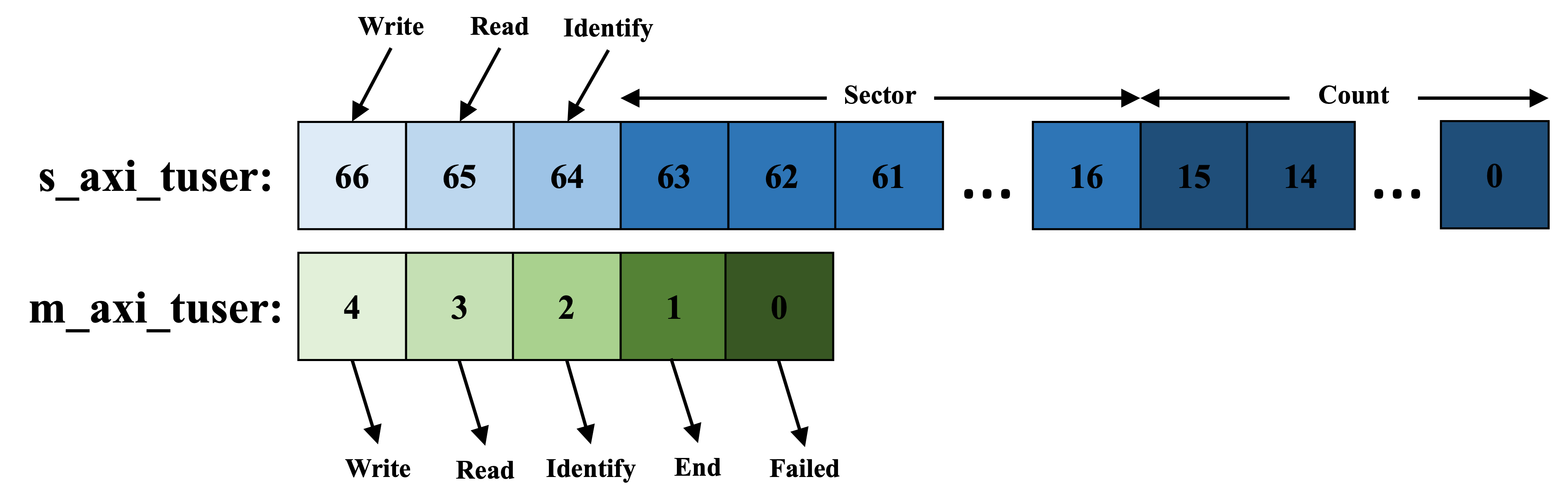}
    \caption{User signal specifications for transceiver and receiver}
    \label{fig: satacore_users}
\end{figure}
The SATAcore design incorporates a system clock operating at either 150MHz or 156.25MHz, depending on the SATA generation, SATA II or SATA III. Fig.~\ref{fig: satacore_users} demonstrates the user signal format for both the transmitter (s\_axi\_tuser) and receiver (m\_axi\_tuser). In the case of \textbf{s\_axi\_tuser}, the \textbf{Count} signifies the number of sectors in a chunk transferred in the burst mode, which can hold up to 64K sectors. The \textbf{Sector} denotes the header sector address of a chunk, while \textbf{Identify} returns the SATA device information. \textbf{Read} and \textbf{Write} bits are used to control the operation of SATAcore. On the other hand, \textbf{m\_axis\_tuser} indicates \textbf{Failed} for errors, \textbf{End} for the end of a sector, and \textbf{Identify}, \textbf{Read}, and \textbf{Write} corresponding to the signals presented in \textbf{s\_axi\_tuser}. These signals play a crucial role in the upcoming quantum gate designs that will control the behaviours of SATA disks.

\subsection{IP Designs for Quantum Gates}
\label{subsec: quantum_ip}
In discussing the design of quantum gates, we focus on implementing the set of universal quantum gates using the tool of Vitis HLS 2022.2.2, as mentioned previously. This section details HLS-based designs of referred quantum gates, utilizing the potential of quantum gates on FPGA and exploiting the characteristics of SATA disks. In describing IP designs, we explain the most representative $H$ gate and highlight the differences between other gate designs, especially their computational logic. 

\begin{figure}[htp]
    \centering
    \includegraphics[width=0.6\linewidth]{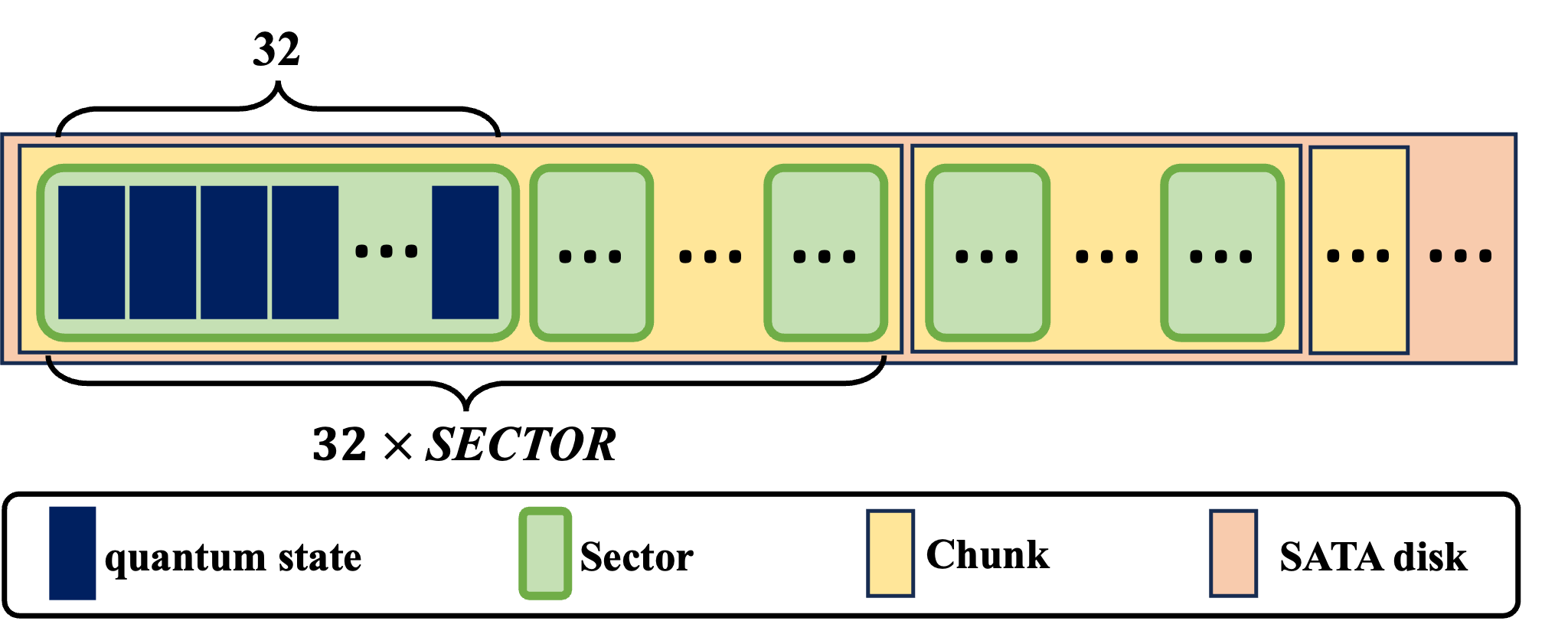}
    \caption{Quantum state allocation in SATA disks}
    \label{fig: memory}
\end{figure}

As illustrated in Fig.~\ref{fig: memory}, in our proposed framework, $CHUNK$ represents the number of SVs in $512\times SECTOR$ bytes. A sector, which is 512 bytes, serves as the communication unit with SATA disks. $SECTOR$ denotes the number of sectors in a single $CHUNK$. Notably, a quantum state, represented by a double-precision floating-point complex number, occupies 16 bytes, allowing 32 quantum states to fit into a single sector. The formula expresses this relationship: $CHUNK = 32 \times SECTOR$.

Furthermore, we present the general parameters outlined in TABLE~\ref{tab: parameter}. These parameters encompass the predefined \textbf{Data} utilized to regulate the burst mode and algorithmic behaviour, as well as \textbf{Input}s and \textbf{Output}s for IP designs. Specifically,  $CHUNK\_INDEX$, in the size of $\log_2{CHUNK}$, governs the behaviour of algorithms referring to a SATA disk; $DIM\_INDEX$, in the size of $\log_2{dim}$, primarily works for the algorithm involving two separate SATA disks, with $dim$ serving as an input parameter indicating the available SVs on a disk.
\begin{table}[htp]
\footnotesize
\vspace*{-0.6em}
\caption{Common parameters settings for quantum gates in a single SATA disk}\label{tab: parameter}
\vspace*{-0.5em}
\centering
    \begin{tabular}{||c|| c | c ||} 
        \hline
        \textbf{Categ.} & \textbf{Parameter} & \textbf{Function} \\
        \hline\hline
        \multirow{4}{*}{\textbf{Data}} & $CHUNK$ & Controlling the burst mode \\
        \cline{2-3}
        & $CHUNK\_INDEX$ & Qubits available in a chunk\\ 
        \cline{2-3}
        & $DIM\_INDEX$ & Qubits available in a SATA disk\\ 
        \cline{2-3}
        & $SECTOR$ & Number of sectors in a chunk\\ 
        \hline
        \multirow{3}{*}{\textbf{Input}} & $dim$ & Available SVs on a disk \\
        \cline{2-3}
        & $target$ & Index of target qubit\\
        \cline{2-3}
        & $qread$ & States saved in SATA disks\\
        \hline
        \textbf{Output} & $qwrite$ & States after the gate processing\\
        \hline
    \end{tabular}
\end{table}

Finally, in the design of quantum gate implementations in the Qulacs library \cite{Qulacs}, gate operations are categorized into two distinct cases depending on whether the $target$  is 0 or not, which is driven by data access optimization. Manipulating the least significant bit in the binary representation of a quantum state ($target = 0$) involves more frequent bit flips and interleaved data processing, resulting in increased computational overhead. To address this, a particular case where $target = 0$ is adopted without negatively impacting the overall system optimization for cases of nonzero target qubit. Following this approach, the quantum gate design in Qu-Trefoil includes specialized modules to efficiently handle the case of $targe = 0$, ensuring optimized performance while managing the added complexity associated with the least significant bit.

\subsubsection{Hadamard Gate}
\label{subsubsection: H_gate}
As a fundamental quantum gate used in quantum computing, the Hadamard gate plays a crucial role in quantum teleportation, error correction, and search algorithms. By applying the Hadamard gate to a single qubit, a user can convert the basis states $\ket{0}$ and $\ket{1}$ into superposition states, which are a combination of both basis states as demonstrated in TABLE~\ref{tab: gates}. Specially:
\begin{equation*}
\begin{aligned}[t]
  H\ket{0} = \frac{1}{\sqrt{2}}(\ket{0} + \ket{1}),
\end{aligned}
\quad\text{and}\quad
\begin{aligned}[t]
 H\ket{1} = \frac{1}{\sqrt{2}}(\ket{0} - \ket{1}).
\end{aligned}
\end{equation*}

Regarding the implementation of the $H$ gate in Qulacs \cite{Qulacs}, the design consists of two cases considering the location of the target qubit: $target$ is 0 or not. However, when it comes to implementing the gate with SATA disks, there are four different situations to account for depending on the target qubit: (1) $target$ being 0 ($H\_0$); (2) referred states within a chunk ($H\_chunk$); (3) referred states across chunks while within a disk ($H\_disk$); (4) referred states across disks ($H\_system$).

Here are the algorithm details for $H\_0$ in Algorithm~\ref{alg: H_0}. The algorithm has three main parts: SVs assignment to buffer in Chunk size (Lines 5-7), computation referring to adjacent states (Lines 8-11), and output to SATA disks (Lines 13-16). Lines 4 and 12 are for the user signal setting, which controls SATA's read/write as introduced in Section~\ref{subsec: SATA_FPGA}. To ensure smooth pipelining, we have intentionally isolated these two parts and designated them as \textbf{\textit{fixed}} protocol regions.
\begin{algorithm}[hbt!]
\footnotesize
\caption{$H\_0$ design in Qu-Trefoil}\label{alg: H_0}
\SetAlgoLined
\uIf{$target$ is 0}{
    \For{$index \gets 0$ \KwTo $dim$} {
        \Comment{Sector of s\_axi\_tuser}
         $sect\_ad \gets index \gg CHUNK\_INDEX$\;
    
        \Comment{\ding{172} Data assignment to buffer}
        $user \gets user\_read(sect\_ad, SECTOR)$\;
        \For{$bufi \gets 0$ \KwTo $CHUNK$} {%
            $state[bufi] \gets COMPLEX(qread[2*bufi], qread[2*bufi+1]$\;
        }
        \Comment{\ding{173} Quantum gate computation}
        \For{$bufi \gets 0$ \KwTo $CHUNK$} {
            $H\_Comp\_0(state)$\;
            $bufi \gets bufi+2$\;
        }
        \Comment{\ding{174} Write result to SATA Disk}
        $user \gets user\_write(sect\_ad, SECTOR)$\;
         \For{$bufi \gets 0$ \KwTo $CHUNK$} {%
            $qwrite[2*bufi] \gets state[bufi].real()$\;
            $qwrite[2*bufi + 1] \gets state[bufi].imag()$\;
        }
        $index \gets index + CHUNK$\;
    }
}
\end{algorithm}

\begin{figure}[htp]
    \centering
    \includegraphics[width=0.8\linewidth]{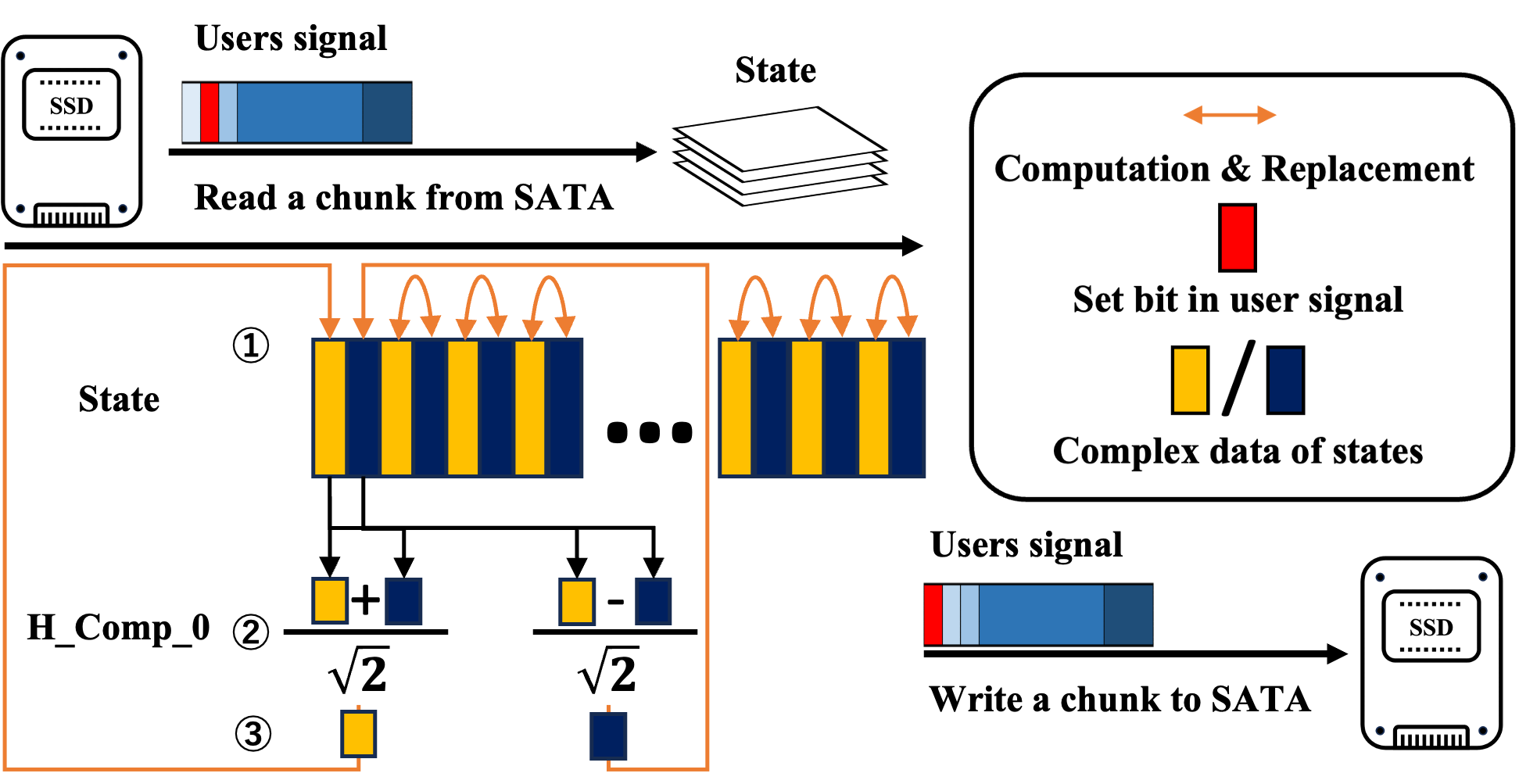}
    \caption{$H\_0$ implementation with pipelining the reading and computation}
    \label{fig: H_0}
\end{figure}

To boost the performance of the IP, we utilize HLS techniques while ensuring the integrity of the algorithm's logic. As presented in Fig.~\ref{fig: H_0}, we explore the on-chip resources by unrolling the $H\_Comp\_0$ in different degrees. Furthermore, considering the SATA disks' inefficient data transfer, we provide two cases to optimize the system's performance: (1) Pipelining the reading part with computation as illustrated in Fig.~\ref{fig: H_0} and (2) Pipelining the writing part with computation, which we will further assess in Section~\ref{subsubsection: H_eval}.

\begin{figure}[htp]
    \centering
    \includegraphics[width=0.9\linewidth]{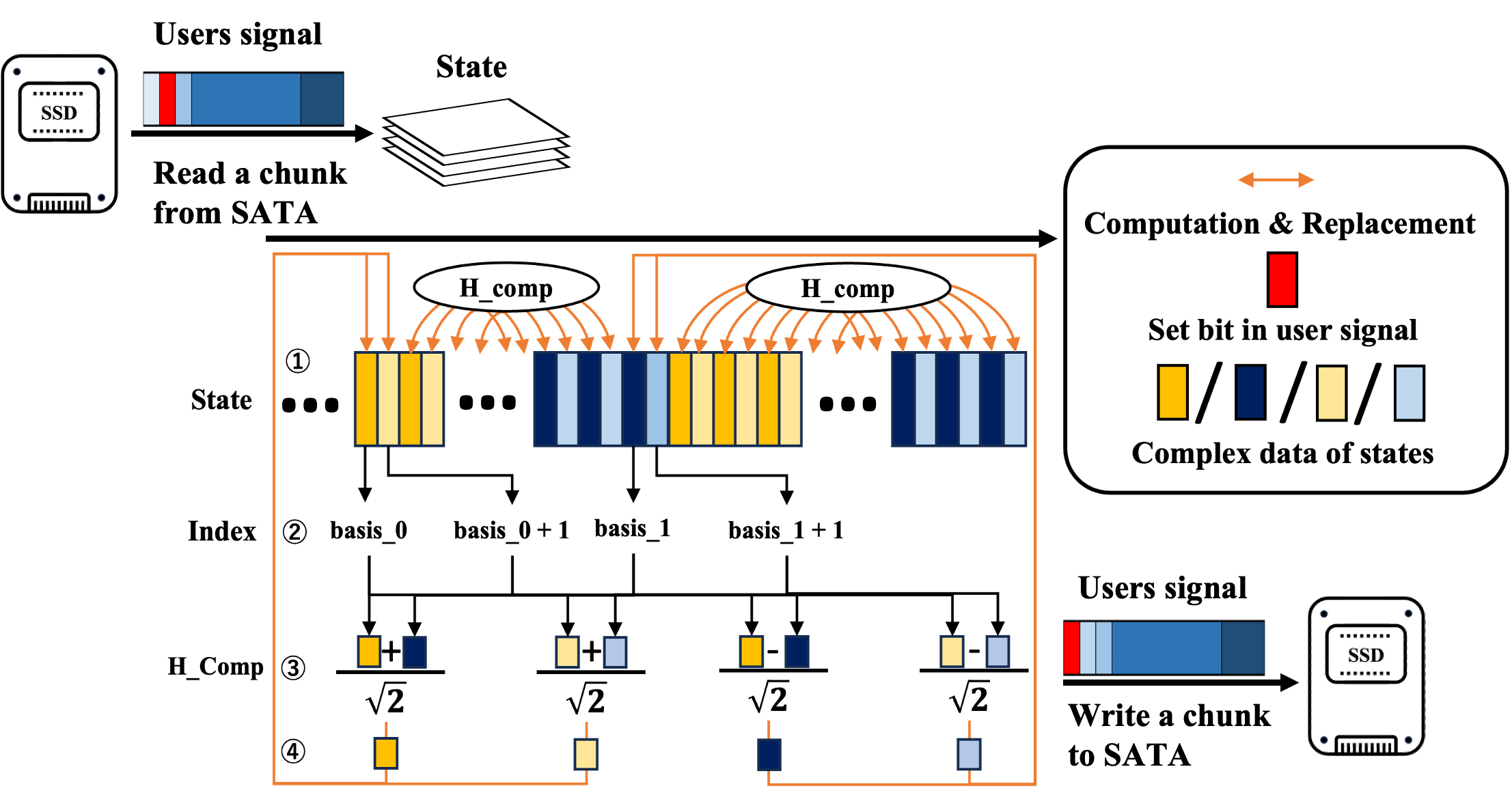}
    \caption{$H\_chunk$ optimization in a single chunk}
    \label{fig: H_small}
\end{figure}

In the case of $H\_chunk$, where $target < CHUNK\_INDEX (\log_2{CHUNK})$, the $H\_comp$ of referred states relates to $mask, low, high$, as defined in Lines 3 and 4 in Algorithm \ref{alg: H_small}. Due to the random data access involved in the $H\_comp$, complete pipeline processing is infeasible for this case, as denoted Fig.~\ref{fig: H_small}. Thus, we emphasize the optimization of the quantum gate computation of pipelining the process \ding{173} and \ding{174} with unrolling the $H\_comp$ thanks to these four independent processes as presented in \ding{174} of Fig.~\ref{fig: H_small} under the constraints of on-chip resources.

\begin{algorithm}[hbt!]
\footnotesize
\caption{$H\_chunk$ design in Qu-Trefoil}\label{alg: H_small}
\SetAlgoLined
\uIf{$target < CHUNK\_INDEX$}{
    $loop\_dim \gets \frac{dim}{2}$\Comment*[r]{Loop constrain}
    $mask \gets 1 \ll target$\;
    $low\_M \gets mask - 1$, $high\_M \gets \sim low\_M$\;
    \For{$index \gets 0$ \KwTo $loop\_dim$} {%
        \Comment{Sector of s\_axi\_tuser}
         $sect\_ad \gets index \gg CHUNK\_INDEX$\;
        \Comment{\ding{172} Data assignment to buffer}
        $user \gets user\_read(sect\_ad, SECTOR)$\;
        \For{$bufi \gets 0$ \KwTo $CHUNK$} {%
            $state[bufi] \gets COMPLEX(qread[2*bufi], qread[2*bufi+1]$\;
        }
        \Comment{Quantum gate computation}
        \For{$bufi \gets 0$ \KwTo $\frac{CHUNK}{2}$} {
	        \Comment{\ding{173} Basis computation}
            $basis\_0 \gets
				(((index + bufi) \ \&\  low\_M) + (((index + bufi) \ \&\  high\_M) \ll 1)) \ \&\  (CHUNK-1)$\;
			$basis\_1 \gets (basis\_0 + mask) \ \&\  (CHUNK-1)$\;
	        \Comment{\ding{174} $H$ gate computation}
            $H\_Comp(state)$\;
            $bufi \gets bufi+2$\;
        }
        \Comment{\ding{175} Write result to SATA Disk}
        $user \gets user\_write(sect\_ad, sector)$\;
         \For{$bufi \gets 0$ \KwTo $CHUNK$} {%
            $qwrite[2*bufi] \gets state[bufi].real()$\;
            $qwrite[2*bufi + 1] \gets state[bufi].imag()$\;
        }
        $index \gets index + CHUNK$\;
    }
}
\end{algorithm}

\begin{figure}[htp]
    \centering
    \includegraphics[width=0.9\linewidth]{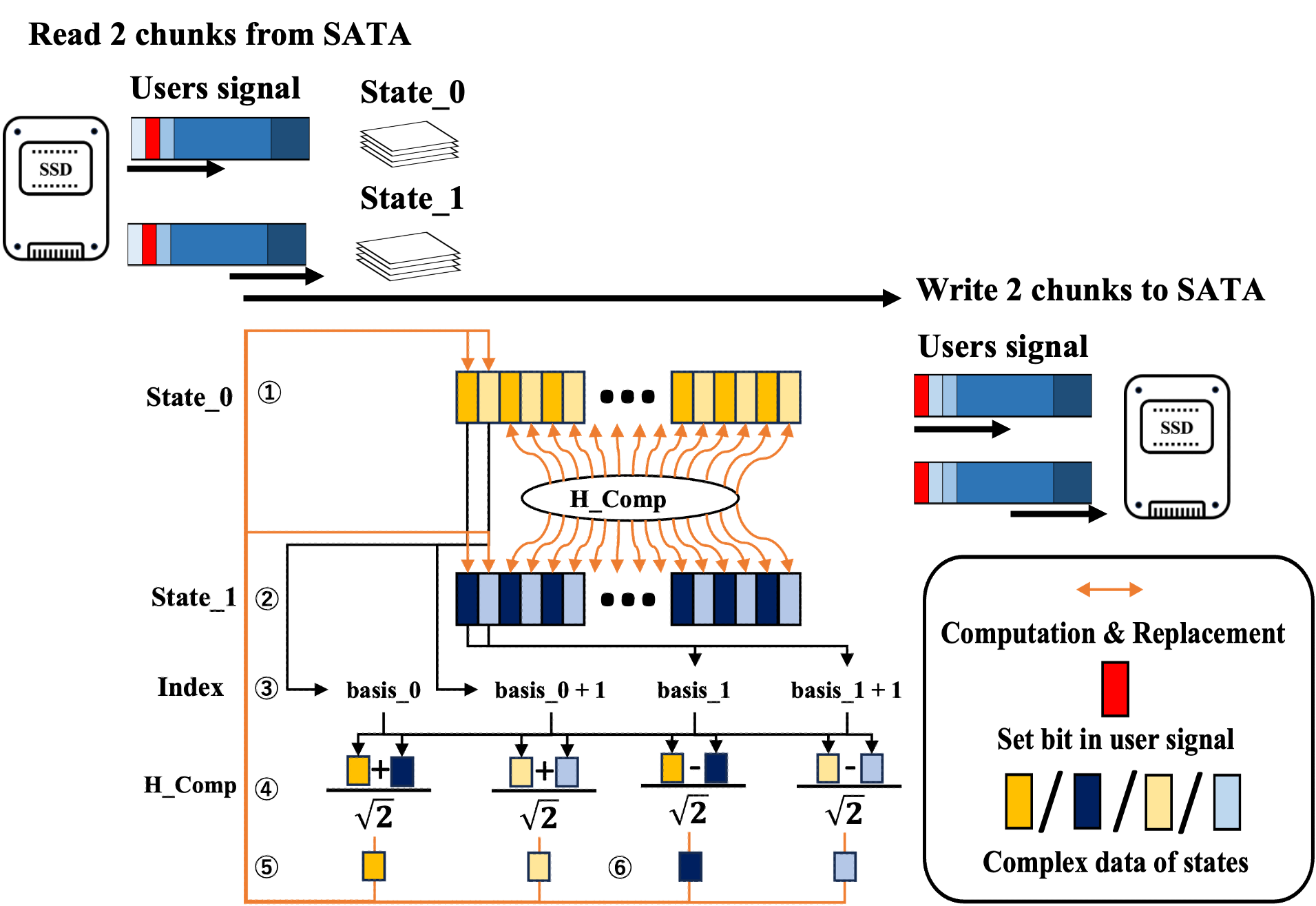}
    \caption{$H\_disk$ design considering two chunks}
    \label{fig: H_large}
\end{figure}

In the case of $H\_disk$, which $CHUNK\_INDEX \leq target < DIM\_INDEX (log_2{dim})$, referred states are included in distinct chunks of a disk. Referencing Fig.~\ref{fig: H_large}, it is necessary to read two chunks from a SATA disk sequentially by setting different $Sector$ in $s\_axi\_tuser$ and load to separated buffers, $State\_0$ and $State\_1$, which is quite different from $H\_chunk$. On the other hand, after the quantum gate computation, the results distributed in these two chunks have to be written back to SATA disks successively. Thus, performance degrades due to the two additional disk accesses required compared to $H\_chunk$.

Finally, in the case of $H\_system$, where the referred two chunks are in two separate disks, we utilize two input/output streams and user signals to accept/output chunks from/to two distinct disks. With the expansion of $N$ SATA disks, the qubits simulation scale increases by $log_2{N}$ qubits. The case of $H\_system$ is to handle the target qubit in the $[DIM\_INDEX, DIM\_INDEX+log_2{N})$. The mechanism of $H\_system$ is quite similar to the design of $H\_disk$. The main difference is the design of the input and output parts. 

\subsubsection{Pauli-Z Gate}
\label{subsubsection: Z_gate}
As another fundamental single-qubit quantum gate, the $Z$ gate rotates the qubit state around the Z-axis of the Bloch sphere, which leaves the state $\ket{0}$ unchanged and introduces a phase of $-1$ to the state $\ket{1}$. Formally:
\begin{equation*}
\begin{aligned}[t]
  Z\ket{0} = \ket{0},
\end{aligned}
\quad\text{and}\quad
\begin{aligned}[t]
 Z\ket{1} = - \ket{1}.
\end{aligned}
\end{equation*}

Regarding HLS design, the $Z$ gate design closely resembles the $H$ gate in four cases. For the case of $target$ is $0$ ($Z\_0$), the computation part (Line 9 of Algorithm~\ref{alg: H_0}) is replaced with $State[bufi]\gets State[bufi]*(-1)$. While for the $Z\_chunk$, the computation part is similarly replaced, but both $State[bufi]\gets State[bufi]*(-1)$ and $State[bufi+1]\gets State[bufi+1]*(-1)$ are applied. Since the $Z$ gate operates on a diagonal unitary matrix, the phase shift is applied to each state independently, making it unnecessary to reference states in distinct chunks during computation. Consequently, in Qu-Trefoil's design, $Z\_disk$ and $Z\_system$ do not exist. The computational complexity of the $Z$ gate is less demanding than that of the $H$ gate, meaning that optimization of the $Z$ gate is more closely tied to on-chip memory resources.

\subsubsection{Phase Gate}
\label{subsubsection: S_gate}
The $S$ gate introduces a $phase$ shift of $\frac{\pi}{2}$ to the state of $\ket{1}$ while leaving the state $\ket{0}$ unchanged, as introduced in TABLE~\ref{tab: gates}. When applied to a qubit, the $S$ gate induces the following effects:
\begin{equation*}
\begin{aligned}[t]
  S\ket{0} = \ket{0},
\end{aligned}
\quad\text{and}\quad
\begin{aligned}[t]
 S\ket{1} = (e^{\frac{i\pi}{2}})\ket{1},
\end{aligned}
\end{equation*}
where $e^{\frac{i\pi}{2}}$ is a complex number corresponding to a $\frac{\pi}{2}$ phase shift, as $e^{\theta} = \cos\theta + i\sin\theta$. For $\theta = \frac{\pi}{2}$, $e^{\frac{i\pi}{2}} = i$.

For the $S$ gate's HLS design, an additional parameter, $phase$, is introduced to the system, distinguishing it from previous designs. When the target is 0, the computation step is modified as $State[bufi] \gets State[bufi] * i$, with the $index$ starting from $1$ instead of $0$ to manage the SV access across all disks. For nonzero targets in the other three cases, the computation is dominated by $State[bufi] \gets State[bufi] * i$ and $State[bufi+1] \gets State[bufi+1] * i$, which closely resembles the nonzero target case of the $Z$ gate. Since the $S$ gate, like the $Z$ gate, operates on diagonal unitary matrices, the $S\_disk$ and $S\_system$ components are not required in the design. Based on the computational logic presented, the independent processing allows for unrolling the computational loop, with the $factor$ controlling resource utilization effectively.

\subsubsection{Controlled-NOT Gate}
\label{subsubsection: CNOT_gate}
The $CNOT$ gate is a fundamental two-qubit gate in quantum computing that performs a conditional operation on the target qubit (the second qubit) based on the state of the control qubit (the first qubit). The $CNOT$ gate flips the state of the target qubit when the control qubit is in the state of $\ket{1}$, which has the following effects:
\begin{equation*}
\begin{aligned}[t]
CNOT\ket{00} = \ket{00}
\end{aligned}
\quad\text{and}\quad
\begin{aligned}[t]
CNOT\ket{01} = \ket{01},
\end{aligned}
\end{equation*}
\begin{equation*}
\begin{aligned}[t]
CNOT\ket{10} = \ket{11}
\end{aligned}
\quad\text{and}\quad
\begin{aligned}[t]
CNOT\ket{11} = \ket{10}.
\end{aligned}
\end{equation*}

Regarding the HLS design of the $CNOT$ gate, in addition to the parameters defined in TABLE~\ref{tab: parameter}, another input $control$ is to indicate the index of the control qubit. Unlike the previous gate implementations, the implementation cases consider the position of both $target$ and $control$ qubits with specifying the $\bm{loop\_dim \gets \frac{dim}{4}}$: (1) $target$ being $0$ ($CNOT\_0$); For $control=0$ while $target\neq 0$, (2) Referred states being available in the same chunk ($target < CHUNK\_INDEX$) ($CNOT\_C0\_chunk$), (3) Referred states in the distinct chunks of a disk ($CHUNK\_INDEX \leq target < DIM\_INDEX$) ($CNOT\_C0\_disk$), (4) Referred states in the distinct chunks from/to two disks ($DIM\_INDEX \leq target < DIM\_INDEX + \log_2{N}$) ($CNOT\_C0\_system$); For nonzero $target$ and $control$, (5) Referred states in an identical chunk ($CNOT\_C1\_chunk$), (6) Referred states in the distinct chunks of a disk  ($CNOT\_C1\_disk$), and (7) Data access of the distinct chunks from/to two disks ($CNOT\_C1\_system$).

Despite the complex condition classifications, the gate designs are similar to the previous designs, especially in data assignment and write-back. In this part, we only depict the algorithm details in our design, which we can generally categorize into three cases: $target = 0$ ($CNOT\_0$), $control = 0$, while $target \neq 0$ ($CNOT\_C0$), and $target \neq 0\ \&\ control \neq 0$ ($CNOT\_C1$). 
\begin{table}[htp]
\footnotesize
\vspace*{-0.6em}
\caption{Masks for $CNOT$ gate}\label{tab: CNOT_mask}
\vspace*{-0.5em}
\centering
    \begin{tabular}{|| c | c || c | c ||} 
        \hline
        $\bm{target\_M}$& $1\ll target$ & {$\bm{min\_M}$} & $1\ll min$\\
        \hline
        {$\bm{control\_M}$} & $1\ll control$ &{$\bm{max\_M}$} & $1\ll max$ \\
        \hline
        {$\bm{low\_M}$} & $min\_M - 1$ &{$\bm{high\_M}$} & $\sim(max\_M-1)$ \\
        \hline
        \multicolumn{2}{||c||}{{$\bm{mid\_M}$}} & \multicolumn{2}{c||}{$(max\_M - 1) \oplus low\_M$}\\
        \hline
    \end{tabular}
\end{table}

Regarding the computation parts, seven masks listed in TABLE~\ref{tab: CNOT_mask} are parameters for the above three cases, that $min$ is the minimum of $target$ and $control$; $max$ is the other. First, for the case of $CNOT\_0$, the referred data indexed by $basis$ is defined as follows,
\begin{equation*}
    \begin{aligned}
    basis \gets{} &((((index + bufi)\ \&\  mid\_M) \ll 1) + \\
                  &(((index + bufi)\ \&\  high\_M) \ll 2) + \\
                  &control\_M) \ \&\  (CHUNK-1),
    \end{aligned}
\end{equation*}
where $index$ is the outmost loop parameter for the iteration of a disk, and $bufi$ is the loop parameter for a chunk. The data swap continuously happens to the neighbouring two states indexed by $basis$ and $basis+1$, which exits strong inter-level data dependency. We can hardly optimize the target IP based on such characteristics.

On the other hand, for the case of $CNOT\_C0$, the data with the index of $basis\_0$ swaps with $basis\_1$, which we present the basis computation as follows,
\begin{equation*}
    \begin{aligned}
    basis\_0 \gets{}&((index + bufi) \ \&\  low\_M + \\
		            &(((index + bufi)\ \&\  mid\_M) \ll 1) + \\
                    &(((index + bufi)\ \&\  high\_M) \ll 2) + \\
                    &control\_M) \ \&\  (CHUNK-1),
    \end{aligned}
\end{equation*}
\begin{equation*}
    basis\_1 \gets (basis\_0 + target\_M)\ \&\  (CHUNK-1)
\end{equation*}.

Finally, for the nonzero $control$ and $target$ ($CNOT\_C1$), the basis computations are the same as the case of $CNOT\_C0$. However, there are four referred data in this case: $basis\_0$, $basis\_1$, $basis\_0+1$, $basis\_1+1$, that the data of $basis\_0$ swaps with $basis\_1$ and $basis\_0+1$ swaps with $basis\_1+1$.

As the data swaps in these four cases operate independently, we can leverage HLS techniques like pipelining or unrolling to improve the $CNOT$ gate computation's performance. Thanks to the $CNOT$ gate not including any intensive computation process with only data swapping, the target IPs are not sensitive to computational resources. Our design should focus solely on tailoring on-chip memory resources.

\subsubsection{T gate}
\label{subsubsection: T}
The T gate, a non-Clifford quantum gate, is a fundamental single-qubit gate that leaves the $\ket{0}$ state unchanged while applying a phase shift of $\frac{\pi}{4}$ to the $\ket{1}$ state:
\begin{equation*}
\begin{aligned}[t]
 T\ket{0} = \ket{0},
\end{aligned}
\quad\text{and}\quad
\begin{aligned}[t]
 T\ket{1} = (e^{\frac{i\pi}{4}})\ket{1}.
\end{aligned}
\end{equation*}
This is similar to the $S$ gate, with the key difference being the amount of phase shift. Since $e^{\frac{i\pi}{4}} = \frac{1+i}{\sqrt{2}}$, the $\frac{\pi}{4}$ phase shift results in the $\ket{1}$ state being multiplied by $\frac{1+i}{\sqrt{2}}$. Since the HLS design for the $T$ gate also operates on diagonal unitary matrices, the design can largely follow the $S$ gate. The main modification involves changing the parameter $phase$ from $i$ to $\frac{1+i}{\sqrt{2}}$. Additionally, the design considers only two cases: $target = 0$ ($T\_0$) and $target \neq 0$ ($T\_chunk$).

\subsubsection{Unitary Matrix}
\label{subsubsection: matrix}
To demonstrate robustness in handling quantum gates with intensive computation, we adopted a general $4 \times 4$ unitary transformation in quantum mechanics, resulting in reversible quantum state transformations. Since this quantum gate operates on two target qubits, we explicitly address the four relevant states concerning chunk and disk. To simplify our implementation, we fixed one of the target qubits to 0, effectively applying the identity gate to qubit 0, while the unitary transformation is applied to the other target one. This design allows conditional operations on one qubit while maintaining the other in a fixed state, thereby managing the intensive computational complexity.

Regarding the implementation, in addition to parameters set in TABLE~\ref{tab: parameter}, there is one more \textbf{Input} named $matrix$, a buffer with 16 complexes accepted as a quantum gate matrix. Since we have simplified our designs, we discuss three cases in the IP design: (1) referred states within a chunk ($Matrix\_chunk$); (2) referred states across chunks while within a disk ($Matrix\_disk$); (3) referred states across disks ($Matrix\_system$).

\begin{figure}[htp]
    \centering
    \includegraphics[width=0.9\linewidth]{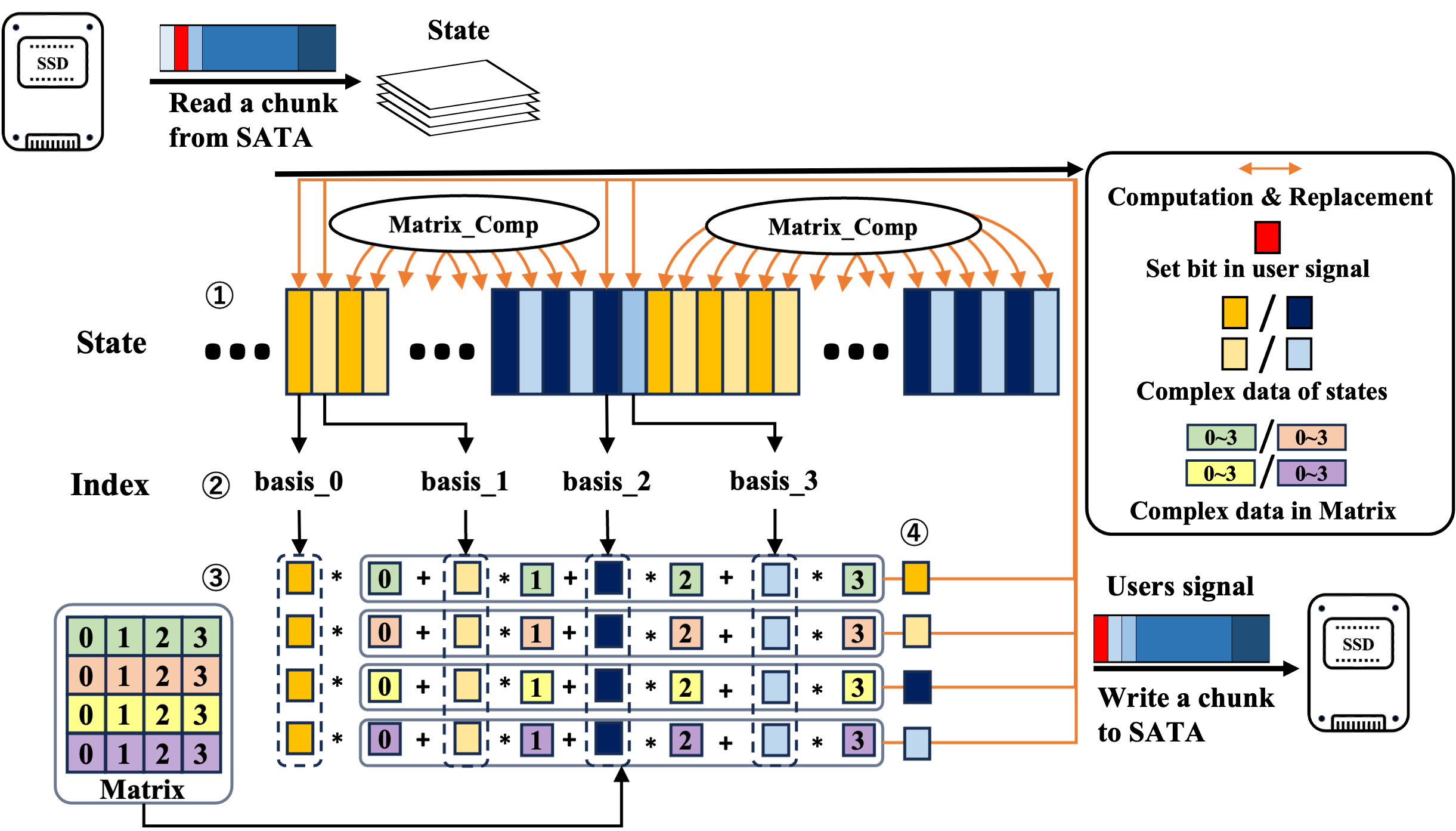}
    \caption{$Matrix\_chunk$ implementation with a single chunk}
    \label{fig: Matrix_small}
\end{figure}

In discussing $Matrix\_chunk$, as presented in Algorithm~\ref{alg: Matrix_small}, since the referred states are in the same chunk, the algorithm consists of a single read/write from/to SATA disk combined with basis and unitary matrix computation, which is similar to $H\_chunk$. Since the intensive complex data computation presented in Fig.~\ref{fig: Matrix_small} is independent, in optimizing the performance of the computation part, we parallelized the \ding{173} and \ding{174} using the techniques of buffering and unrolling under the constraints of on-chip memory.
\begin{algorithm}[hbt!]
\footnotesize
\caption{$Matrix\_chunk$ design in Qu-Trefoil}\label{alg: Matrix_small}
\SetAlgoLined
\uIf{$target < CHUNK\_INDEX$}{
    \Comment{Mask assignments}
    $min\_index \gets 0$\Comment*[r]{Fixing a $target$ to 0}
    $max\_index \gets target$, $min\_M \gets 1 \ll min\_index$\;
    $max\_M\gets 1 \ll (max\_index-1)$, $low\_M\gets min\_M - 1$\;
    $mid\_M\gets (max\_M - 1)\oplus low\_M$, $high\_M\gets \sim(max\_M - 1)$\;
    $mask\_1\gets 1\ll 0$, $mask\_2\gets 1\ll traget$\;
    $loop\_dim \gets \frac{dim}{4}$\Comment*[r]{Loop constrain}
    \For{$index \gets 0$ \KwTo $loop\_dim$} {%
        \Comment{Sector of s\_axi\_tuser}
        $basis\_0 \gets (index \ \&\  low) + (index \ \&\  mid) \ll 1 +
        (index \ \&\  high) \ll 2 $\Comment*[r]{Create index}
        $sect\_ad \gets basis\_0 \gg CHUNK\_INDEX$\;
        \Comment{\ding{172} Data assignment to buffer}
        $user \gets user\_read(sect\_ad, SECTOR)$\;
        \For{$bufi \gets 0$ \KwTo $CHUNK$} {%
            $state[bufi] \gets COMPLEX(qread[2*bufi], qread[2*bufi+1]$\;
        }
        \Comment{Quantum gate computation}
        \For{$bufi \gets 0$ \KwTo $\frac{CHUNK}{4}$} {
	        \Comment{\ding{173} Basis computation}
            $basis\_0 \gets
				(((index + bufi) \ \&\  low) + ((index + bufi) \ \&\  mid) \ll 1 + (index +bufi) \ \&\  high) \ll 2 \ \&\  (CHUNK-1)$\;
			$basis\_1 \gets (basis\_0 + mask\_1) \ \&\  (CHUNK-1)$\;
            $basis\_2 \gets (basis\_0 + mask\_2) \ \&\  (CHUNK-1)$\;
            $basis\_3 \gets (basis\_1 + mask\_2) \ \&\  (CHUNK-1)$\;
	        \Comment{\ding{174} $Matrix$ computation}
            $Matrix\_Comp(matrix, state)$\;
        }
        \Comment{\ding{175} Write result to SATA Disk}
        $user \gets user\_write(sect\_ad, sector)$\;
         \For{$bufi \gets 0$ \KwTo $CHUNK$} {%
            $qwrite[2*bufi] \gets state[bufi].real()$\;
            $qwrite[2*bufi + 1] \gets state[bufi].imag()$\;
        }
        $index \gets index + CHUNK$\;
    }
}
\end{algorithm}

Concerning the case of $Matrix\_disk$, referred states exist in two chunks in a disk, similar to the case of $H\_disk$ illustrated in Fig.~\ref{fig: H_large}. Due to the two times we read/write, we have to prepare two buffers to save two chunks, which severely burdens computation and storage resources, and we will further evaluate in Section \ref{sec: eval}.

Finally, for the design of $Matrix\_system$, whose $target$ lays in the range of $[DIM\_INDEX, DIM\_INDEX+log_2{N})$ as discussed in the case of $H\_system$. $State\_0[basis\_0]$ and $State\_0[basis\_1]$ read/write chunk from/to one of the disks and $State\_1[basis\_2]$ and $State\_1[basis\_3]$ from/to another with two times setting of the user signal for read/write. Same as the optimization of $Matrix\_disk$, the implementation challenges the on-chip resources.

\subsection{Overall System Implementation}
\label{subsec: overall_sys}
In this section, we introduce the overall system implementation, emphasizing the block design in Vivado 2022.2.2 and inspecting the details of designed IPs in the proposed simulator, Qu-Trefoil. We divide our introduction into two parts: Quantum gate IP referring to a single SATA disk, such as $*\_0$, $*\_chunk$ and $*\_disk$; and IPs referring to two disks, such as $*\_system$. To support a Trefoil storage subsystem with 32 SATA disks ($N=32$), we must create an instance for each SATA disk to ensure that the resource utilization of a quantum gate IP is kept at around $3\%$ when the system runs in full gear. 

\begin{figure}[htp]
    \centering
    \includegraphics[width=0.9\linewidth]{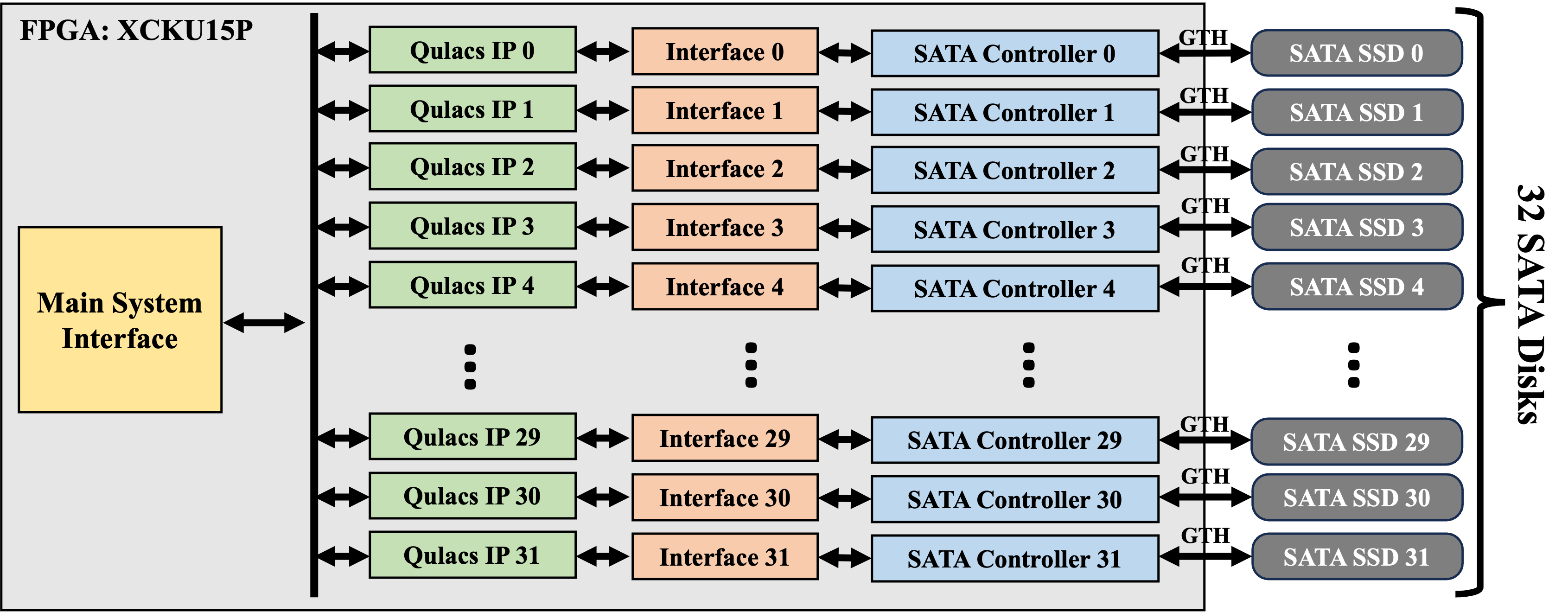}
    \caption{Qu-Trefoil system design of single disk access}
    \label{fig: qutrefoil_single}
\end{figure}

In Fig.~\ref{fig: qutrefoil_single}, we present an overview of Qu-Trefoil's system design, which accesses a single disk per quantum gate IP. The design includes the \textbf{Qulacs IP}s introduced in Section~\ref{subsec: quantum_ip}, \textbf{Interface}s, and \textbf{SATA controller}s. The \textbf{Interface} handles communication between the Qulacs IP and the SATA controller. And \textbf{SATA controller} whose Verilog is generated by LiteX, manages multi-SATA access, as demonstrated in Section~\ref{subsec: SATA_FPGA}. Meanwhile, as shown in Fig.~\ref{fig: trefoil_storage}, the Trefoil storage subsystem connects to a Raspberry Pi 3 via GPIO, functioning as the processing system (PS) for user communication. Since the Raspberry Pi 3 only accepts 32-bit data, the \textbf{Interface} plays a crucial role in separating and concatenating bits, especially when dealing with 64-bit or larger data sizes.

\begin{figure}[htp]
    \centering
    \includegraphics[width=0.9\linewidth]{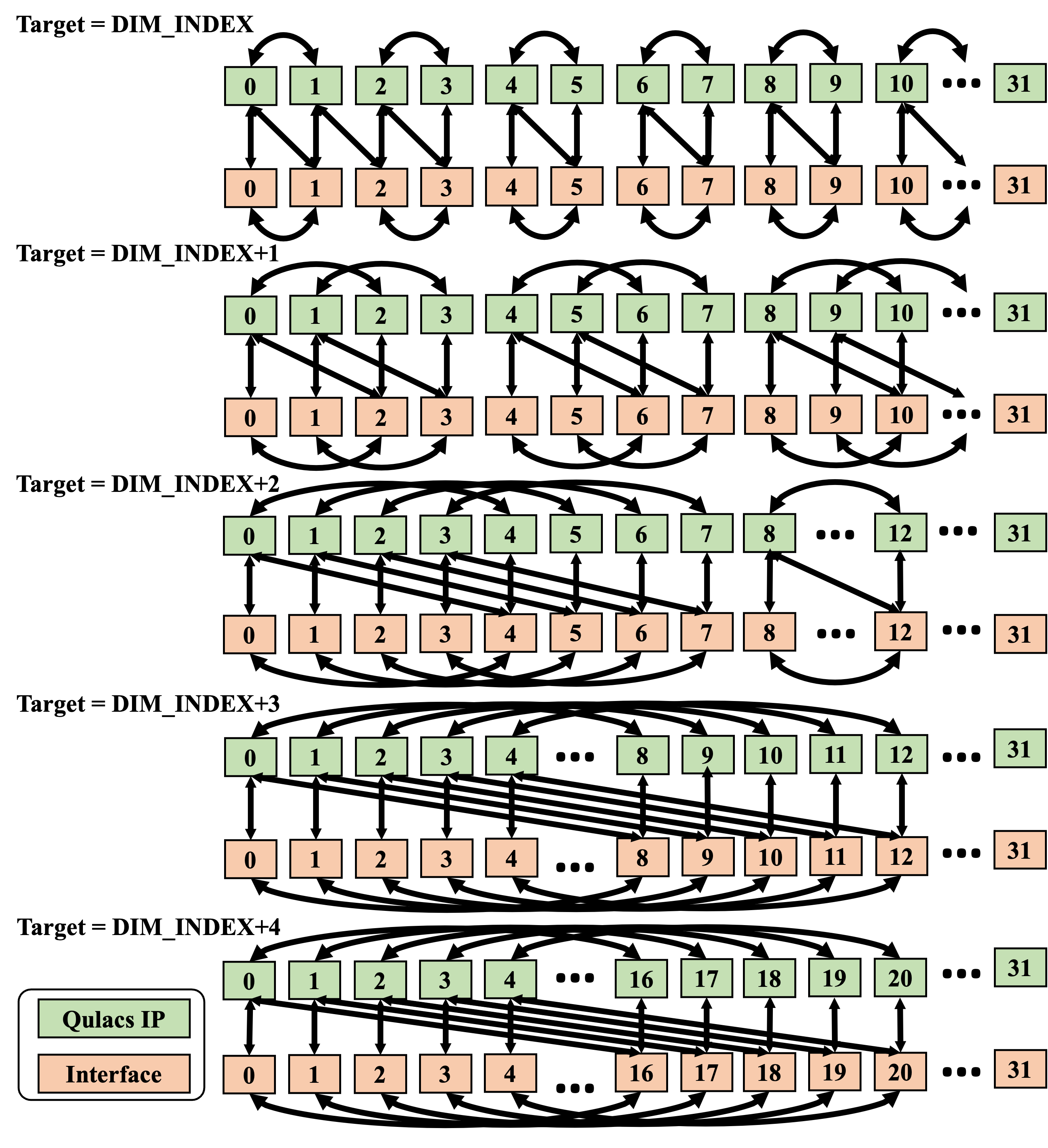}
    \caption{Qu-Trefoil system design for two disks access}
    \label{fig: qutrefoil_across}
\end{figure}

Regarding the case of system design for a Qulacs IP accessing two disks, compared with the previous design presented in Fig.~\ref{fig: qutrefoil_single}, data communications between Qulacs IPs and Interfaces are necessary for quantum gate computation when referred states exist in two chunks from two disks as shown in Fig.~\ref{fig: qutrefoil_across}. Considering 32 disks of a storage subsystem, there are five cases discussed in our design: disk accesses at the interval of $1$, $2$, $4$, $8$ and $16$, corresponding to the $target$ being $DIM\_INDEX$, $DIM\_INDEX+1$, $DIM\_INDEX+2$, $DIM\_INDEX+3$, $DIM\_INDEX+4$, respectively. Since these 32 SATA disks can be accessed simultaneously, as introduced in Section~\ref{sec: trefoil}, there is no variation in both performances and resource utilization for these five cases.

%% file: Contents/5_eval.tex
\section{Evaluation}
\label{sec: eval}
The Qu-Trefoil utilizes Xilinx Kintex Ultrascale+ XCKU15P; its available resources are in TABLE~\ref{tab: trefoil_res}. We have provided the resource utilization of each quantum gate designed in Section~\ref{subsec: quantum_ip}, based on the reports generated by Vivado 2022.2.2. We conducted all measurements on a physical machine to ensure accurate results for simulation speed. In the following sections, we will evaluate Qu-Trefoil's performance with different quantum gate IPs, explore the impact of chunk size and SATA disk extension, and assess the effects of SATA II and SATA III. Furthermore, we will examine the Qu-Trefoil's performance on large qubit scales. Finally, we will compare our system with other top-tier designs based on platform, performance, and costs.

\begin{table}[htp]
\footnotesize
\vspace*{-0.6em}
\caption{Available resources on Trefoil storage subsystem}\label{tab: trefoil_res}
\vspace*{-0.5em}
\centering
    \begin{tabular}{|| c | c || c | c ||} 
        \hline
        $\bm{LUT (K)}$ & $523$ & {$\bm{LUTRAM (K)}$} & $161$\\
        \hline
        {$\bm{FF (K)}$} & $1045$ & {$\bm{BRAM}$} & $984$ \\
        \hline
        {$\bm{DSP}$} & $1968$ & {$\bm{GT}$} & $56$ \\
        \hline
    \end{tabular}
\end{table}

\subsection{Quantum Gates Evaluation}
\label{sec: gate_eval}
\begin{figure*}[htp]
\centering
  \includegraphics[width=0.9\textwidth]{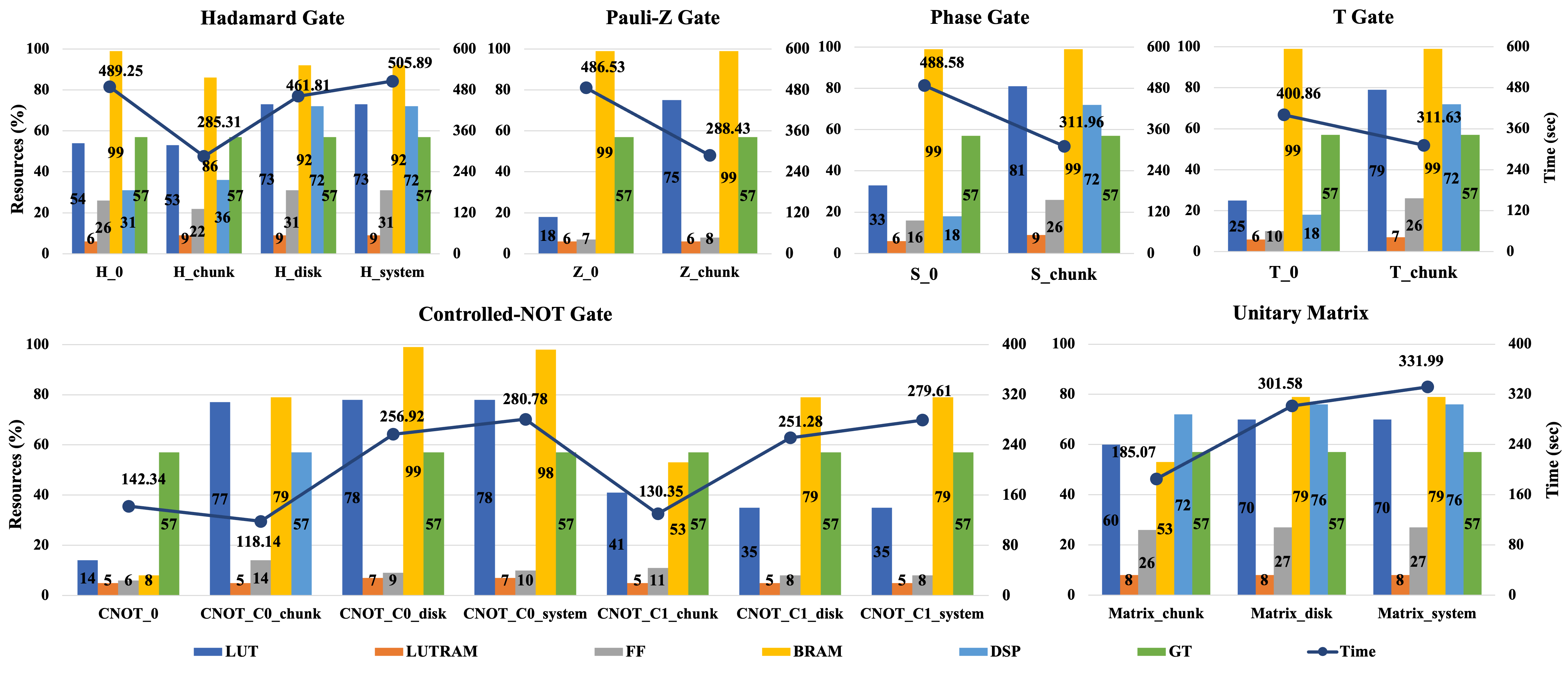}
  \caption{Qu-Trefoil overall systems' evaluation on resources and speed with simulating 35 qubits when SATA disks working at SATA II}
  \label{fig: eval_gates}
\end{figure*}
Regarding the performance of quantum gates in Qu-Trefoil, we evaluate our designs as detailed in Section~\ref{sec: impl}, where a single quantum gate is applied to one qubit, impacting all quantum states within the system. We configure the system with $CHUNK = 512$ at SATA II, which supports 16 sectors in burst mode. Considering both system evaluation efficiency and the upper bound of HPC capabilities, we conduct a 35-qubit simulation utilizing the fully loaded Trefoil storage subsystem connected to 32 SATA disks. The resource utilization and time consumption are illustrated in Fig.~\ref{fig: eval_gates}, where we fully leverage the storage subsystem's capacity to achieve optimal performance.

\subsubsection{Hadmard gate evaluation}
\label{subsubsection: H_eval}
Our evaluation of the $H$ gate includes four cases, as discussed in Section \ref{subsubsection: H_gate}. Due to the pipeline of reading and computation in $H\_0$, the time consumption is mainly for the data communication between SATA disks and FPGA. While $H\_chunk$, $H\_disk$, and $H\_system$ share the same algorithm for $H$ gate computation ($H\_comp$), the difference in their referred data sources results in a performance discrepancy.

The $H\_chunk$ case has only a single chunk referred in the out-most loop count of $\frac{dim}{2}$, as indicated in Algorithm~\ref{alg: H_small}, making it the fastest among the four cases. It achieves almost twice the speed of $H\_0$. However, $H\_disk$ and $H\_system$ refer to two distinct chunks requiring two read/write operations from/to SATA disks, resulting in system performance degradation. To address this, we parallelized the computations in these two chunks using two buffers to accept states, which led to an evident increase in DSP utilization compared to $H\_chunk$. Additionally, due to the time-consuming setup process of SATA disks (requesting data $\Rightarrow$ starting data transfer), $H\_system$ requires a slightly longer time for separate disk accesses than $H\_disk$.

\begin{figure}[htp]
    \centering
    \includegraphics[width=0.8\linewidth]{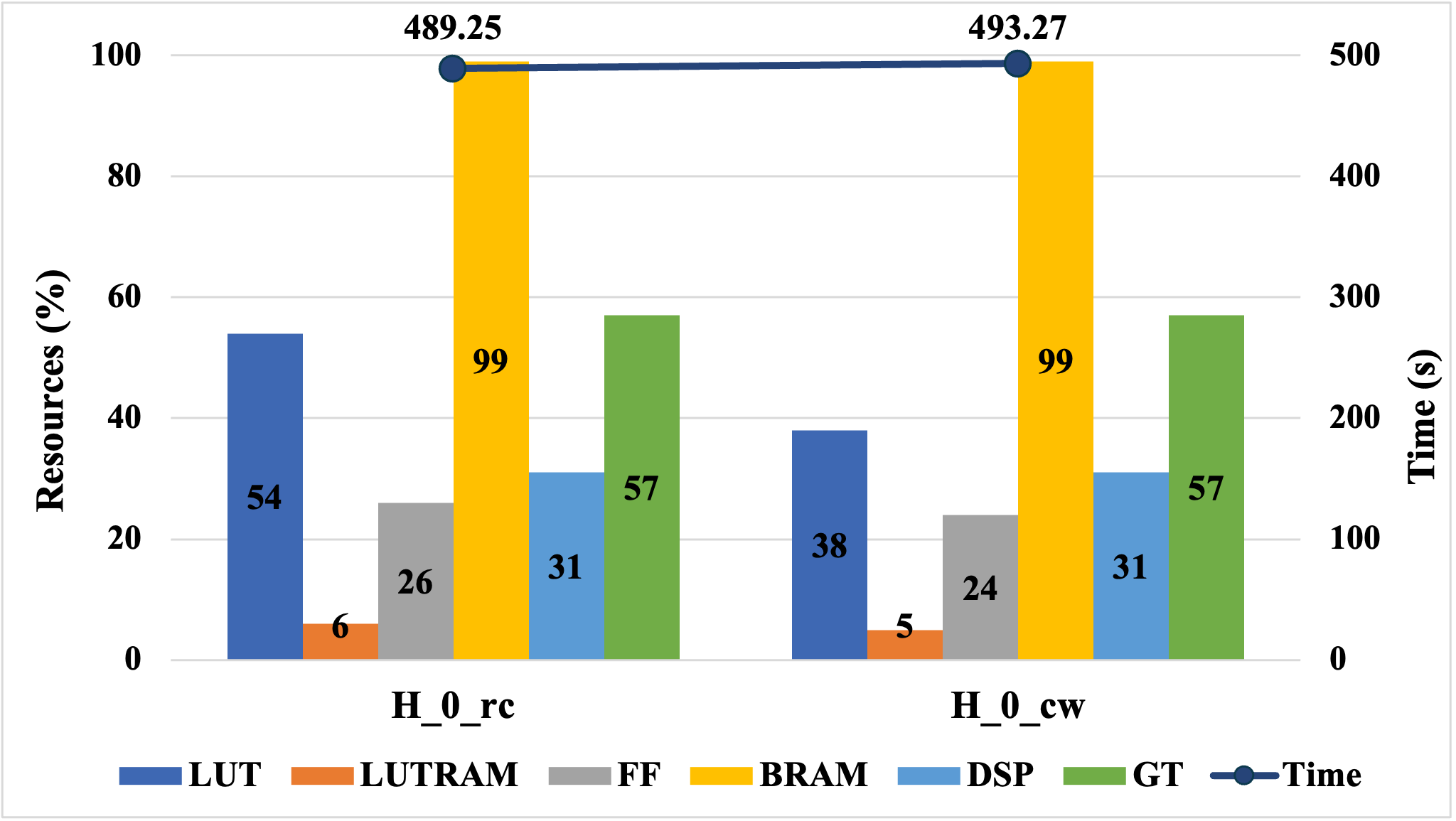}
    \caption{Evaluations upon two strategies for optimizing 35-qubit $H\_0$ implementation with pipelining}
    \label{fig: H_0_comp}
\end{figure}

As mentioned in Section~\ref{subsubsection: H_gate}, we propose two mechanisms for accelerating $H\_0$: pipelining the computation part with reading ($H\_0\_rc$) or writing ($H\_0\_cw$). As shown in Fig.~\ref{fig: H_0_comp}, the two proposed strategies do not make a significant difference. According to the report from Vivado 2022.2.2, $H\_0\_cw$ results in a slight decrease in LUT utilization, which marginally decreases the simulation speed. Thus, the upcoming evaluation will adopt the complete pipeline design of reading and computation.

According to the $H$ gate's on-chip resource utilization, the implementations put tremendous pressure on memory and computation resources, which is quite representative of the overall system evaluation. It costs around $285\sim 506$ sec to realize a 35-qubit $H$ gate simulation.

\subsubsection{Pauli-Z gate evaluation}
As introduced in Section~\ref{subsubsection: Z_gate}, only two cases are included in our design: $Z\_0$ and $Z\_chunk$. Since there is no intensive computation in the algorithm of $Z\_0$, computation resource DSP utilization keeps $0\%$ in the design, according to the Vivado report. As presented in the bar graph, we fully utilize Block RAM (BRAM) to optimize the $Z$ gate logic processing. Thanks to the pipeline processing of reading and computation, we can see that the simulation time is only for data transfer between SATA disks and FPGA. Thus, for a 35-qubit simulation, $Z\_0$ keeps almost the same speed with $H\_0$.

On the other hand, regarding $Z\_chunk$, the running time of $Z\_chunk$ is almost half of the $Z\_0$ because of the decrease in loop iteration. According to our evaluation, we can simulate a 35-qubit $Z$ gate around $288\sim 487$ sec.

\subsubsection{Phase gate evaluation}
Similar to the $Z$ gate design, only two IPs are implemented in Qu-Trefoil, with the reasoning explained in Section~\ref{subsubsection: Z_gate}. For the case of $S\_0$, we have successfully maintained DSP utilization at around $18\%$ by fully pipelining the reading and computation process. For the nonzero target qubit, $S\_chunk$, we partially unroll the quantum gate computation to maximize the use of computational resources (DSP) for performance optimization. As a result, the time required for simulating a 35-qubit $S$ gate is approximately $315\sim 489$ sec.

\subsubsection{Controlled-NOT gate evaluation}
As the most complicated quantum gate in Qu-Trefoil, we have to discuss seven cases in our design, which we introduced in Section~\ref{subsubsection: CNOT_gate}. Since the $CNOT$ gate computation part works on quantum state swapping, the designed IP does not consume any DSP resource. Besides, as the 2-qubit gate, the overall loop iteration is decreased to $frac{dim}{4}$. Thus, there is a dramatic decrease in time consumption compared with the above three quantum gates. For $CNOT\_0$, the data dependency in the algorithm hinders our optimization, so there is no significant resource consumption in our design. In discussing the three cases of $CNOT\_C0$ and $CNOT\_C1$, the performance varies according to the locations of referred states evaluated in Section~\ref{subsubsection: H_eval}. The on-chip BRAM resources eventually restrict the optimization of the $CNOT\_C0$ and $CNOT\_C1$ cases, which we organically pipeline the computation and partially unroll the loop to boost the performance. It consumes around $118\sim 280$ sec to simulate a $CNOT$ gate with 35 qubits.

\subsubsection{T gate evaluation}
As described in Section~\ref{subsubsection: T}, the main difference between the implementations of the $S$ and $T$ gates lies in the parameter setting for the phase shift, changing from $i$ to $\frac{1+i}{\sqrt{2}}$. As expected, this modification does not significantly affect resource allocation, with only a slight decrease in LUT and LUTRAM usage for both cases: $T\_0$ and $T\_chunk$. Regarding processing speed, we observe that the $T$ gate operates slightly faster than the $S$ gate, with a time consumption of approximately $311\sim 400$ sec.

\subsubsection{Unitary matrix evaluation}
The IPs of $4\times 4$ unitary matrix multiplication, shown to be the most computationally-intensive design in Qu-Trefoil, discuss three cases to show the capacity of our target platform, as we introduced in Section~\ref{subsubsection: matrix}. As expected, four intensive double-precision complex data computations hinder further optimization, so we pipeline the computation part to boost the performance. According to the bar graph of the Unitary Matrix, we can notice an exhaustive utilization of overall resources. Eventually, we can simulate the result of 35 qubits in $185\sim 332$ sec.

To summarize, for the $*\_0$'s single-qubit quantum gate designs, we achieve almost the same simulation speed in all considered quantum gates since we pipeline the reading and computation parts, taking advantage of the algorithms' features. Furthermore, for the $CHUNK = 512$ at SATA II, we can assert that $*\_chunk < *\_disk < *\_system$ in time consumption. The $*\_chunk$'s one-time chunk read/write makes its performance exceptional in simulation speed. Although both $*\_disk$ and $*\_system$ have to read two chunks from disks, the time consumption of data requests on separate disks degrades the performance of $*\_system$ at the same resource consumption on the target platform. Regarding two-qubit quantum gate designs ($CNOT$ and $Matrix$), the decrease in loop iteration results in a swift simulation, as presented in Fig.~\ref{fig: eval_gates}.

\subsection{Chunk Size Extension in Burst Mode}
\label{subsec: eval_burst}
\begin{figure*}[htp]
\centering
  \includegraphics[width=0.9\textwidth]{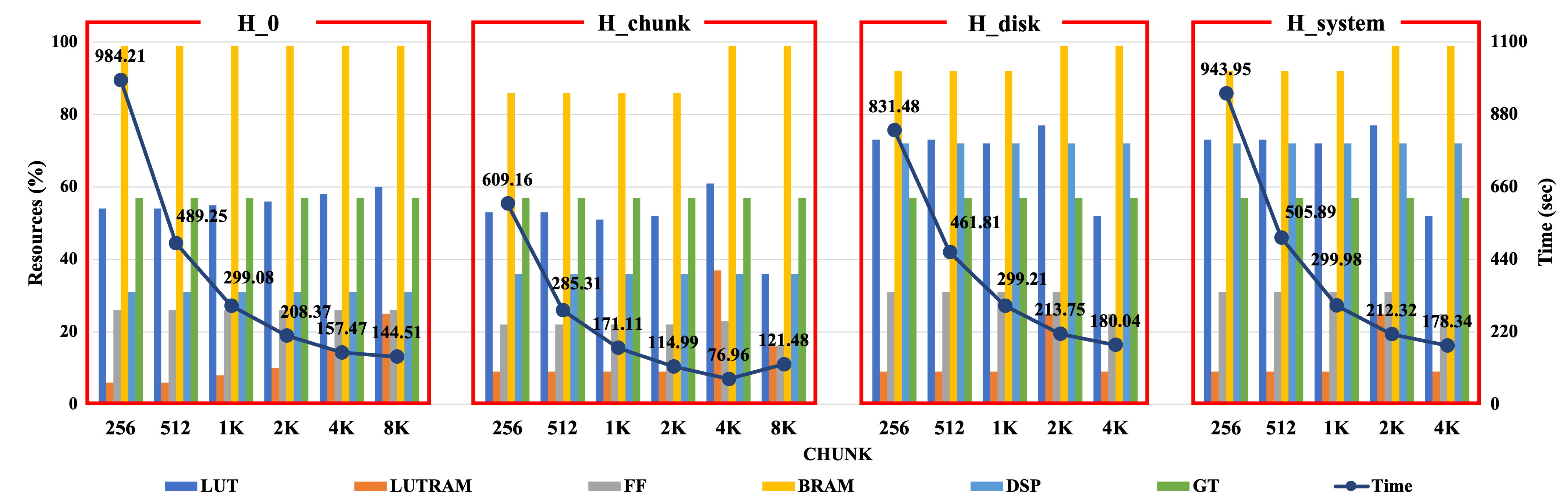}
  \caption{Chunks extension from 256 to 8192 in burst mode working on 35-qubit $H$ gate at SATA II}
  \label{fig: eval_chunks}
\end{figure*}
We adopt the $H$ gate as our target to comprehensively evaluate the burst mode performance of Qu-Trefoil. We thoroughly use on-chip resources cooperating with the $CHUNK = 256, 512, 1024, 2048, 4096, 8192$, corresponding to $8, 16, 32, 64, 128, 256$ sectors per transfer. As in previous quantum gates evaluations, we evaluate the burst mode on resource utilization and speed, as presented in Fig.~\ref{fig: eval_chunks}. With the increase of $CHUNK$ size, the time decreases dramatically in the four cases. Due to the pipeline processing of the reading and computation parts in $H\_0$, we did not notice any significant increase in resource utilization. However, for the other three cases, the optimization mainly works on the computation part, which requires more BRAM to buffer the states with the increase of $CHUNCK$ size. Especially the $H\_chunk$ with the $CHUNK=8192$, there is a slight increase in time consumption compared with the case of $2048$ and $4096$, while an evident decrease in computation resources such as lookup table (LUT), which is because we have to sacrifice the performance for more memory space to fit the design of the target platform. For $H\_disk$ and $H\_system$, we capped $CHUNK$ size at 4096 because of the deficiency of memory space on the Trefoil storage subsystem.

Specifically, according to Fig.~\ref{fig: eval_chunks}, the sloop becomes more and more gentle with the increase of $CHUNK$ size, which means the less effect of disk access speed to Qu-trefoil. In other words, the time consumption of computation gradually dominates the simulation, which further provides evidence for the inconsistent results of $H\_chunk$ with the $CHUNK=8192$. On the other hand, with the increase of $CHUNK$ size, the time consumption of setting up a SATA disk for data transfer becomes less and less, which results in an ideal result that $ H\_system$'s speed is close to that of $H\_disk$.

\subsection{Disks Expansion for Flexibility}
\begin{figure*}[htp]
\centering
  \includegraphics[width=0.9\textwidth]{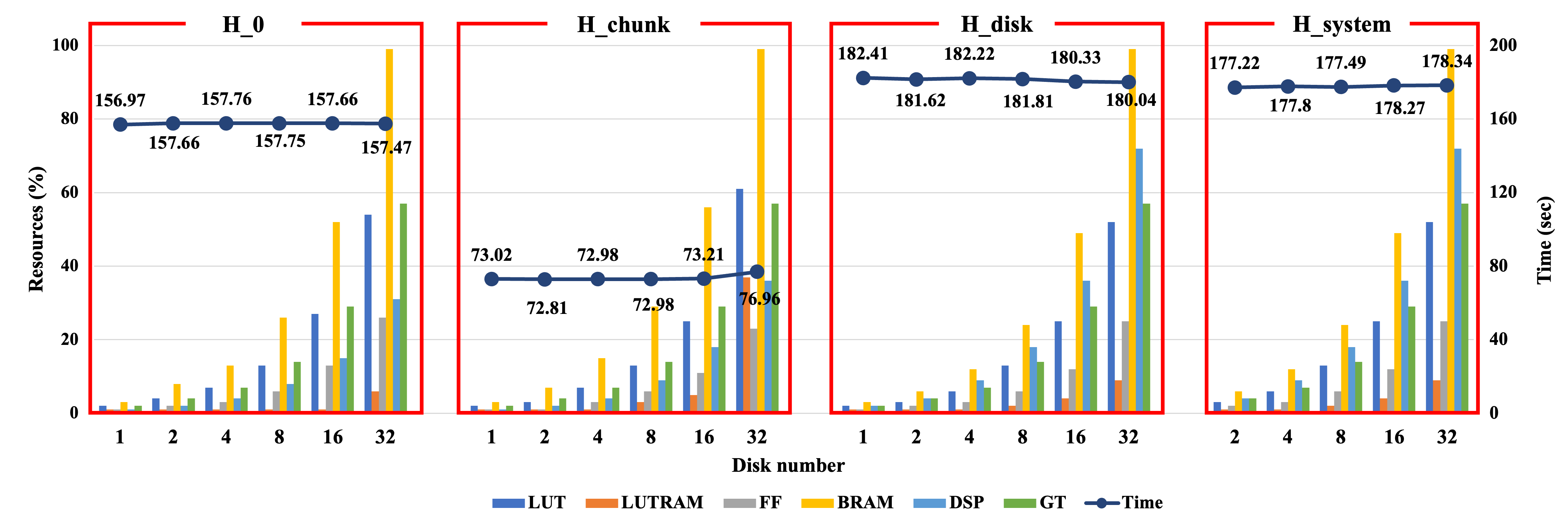}
  \caption{$H$ gate evaluation conducted with disk expansion ranging from 1 to 32 disks, simulating qubit scales from 30 to 35 at SATA II with $CHUNK = 4096$}
  \label{fig: eval_disk}
\end{figure*}
In this section, we evaluate disk expansion in Qu-Trefoil to show the flexibility and robustness of the proposed system. We adopt the H gate simulation, working on the $CHUNCK$ size of 4096 for a fair evaluation. With increased SATA disks, we can simulate more qubits thanks to parallel processing cooperating with multiple SATA disks. Considering the exponential increase of SVs with qubit scales, the SATA disks for SV storage must also increase exponentially to correspond to such a nature. Precisely, $2^n$ disk corresponds to $base + n$ qubits, where $base$ is the number of qubits simulated on one disk. As presented in Fig.~\ref{fig: eval_disk}, with the increase of SATA disks, we can simulate more qubits without any cost in time consumption.

Regarding resource utilization, one Qulacs IP corresponds to one SATA disk for parallel processing on Qu-Trefoil, as introduced in Section~\ref{subsec: overall_sys}. The one-chip resource utilization increases exponentially with more SATA disks involved. From these results, we ascertain the flexibility and robustness of Qu-Trefoil targets on different budgets.

\subsection{Comparison of SATA II and SATA III on Qu-Trefoil}
\label{subsec: eval_gen}
As introduced in Section~\ref{subsec: SATA_FPGA}, the proposed system supports both SATA II and SATA II, which operate at the system frequency of 150MHz and 156.25MHz, respectively. In this section, we keep using the $H$ gate as our target quantum gate at the $CHUNK$ size of 4096. According to the reports from Vivado 2022.2.2, there is no variation in resource utilization for these two generations. Furthermore, due to the speedup of SATA III, we notice a noticeable decrease in time consumption for a 35-qubit simulation.
\begin{figure}[htp]
    \centering
    \includegraphics[width=0.9\linewidth]{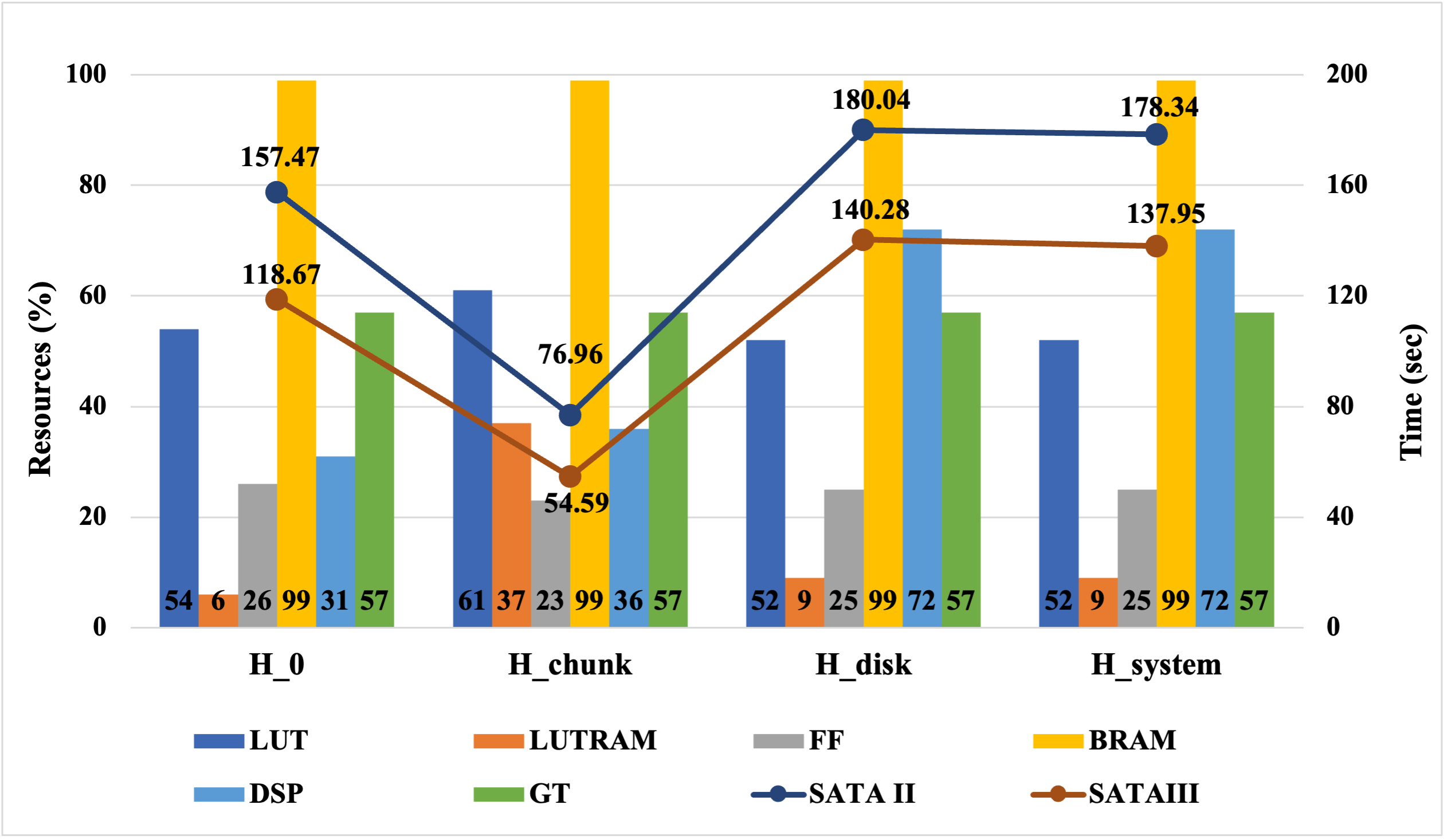}
    \caption{35-qubit $H$ gate simulation conducted with $CHUNK = 4096$ at SATA II and SATA III}
    \label{fig: eval_gen}
\end{figure}

However, according to the specifications of SATA II and SATA III, these interfaces support up to 300MB/s and 600MB/s, respectively. According to Fig.~\ref{fig: eval_gen}, $H\_0$, $H\_chunk$, $H\_disk$ and $H\_system$ achieve the speedup of $24.64\%$, $29.07\%$, $22.08\%$, and $22.64\%$, respectively. To investigate the effects of SATA II and SATA III on Qu-Trefoil's operations, we target $H\_0$ gate IPs for our evaluation, containing just one-time chunk read and chunk write for an iteration. For an appropriate evaluation of SATA II and SATA III, we cancel the optimizations on the algorithm to completely isolate the processes of reading and computation.

\begin{figure}[htp]
    \centering
    \includegraphics[width=0.8\linewidth]{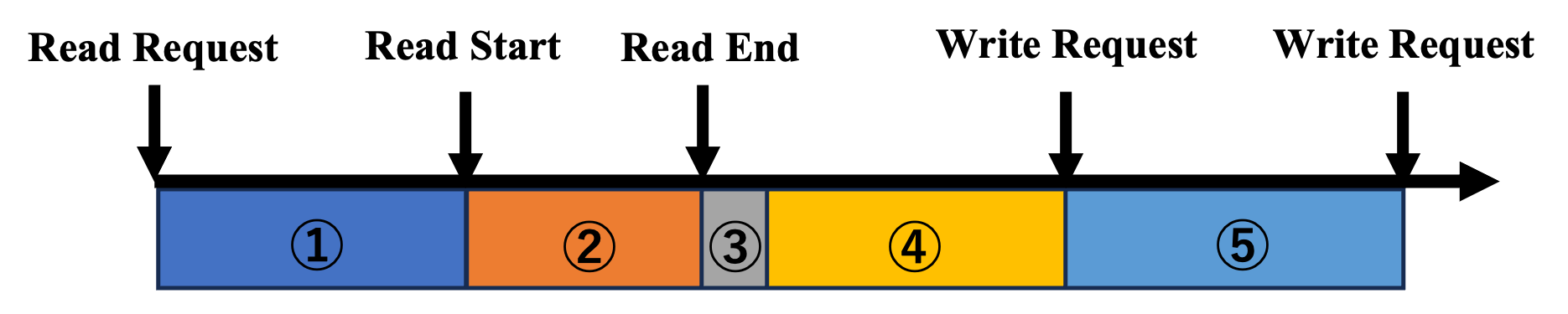}
    \caption{Scheduler for Qu-Trefoil without optimization}
    \label{fig: gen_timeline}
\end{figure}

According to the timers set in our design, the time consumption consists of five steps as presented in Fig~\ref{fig: gen_timeline}:
\begin{dingautolist}{192}
	\item Request for the SV from SATA disks and wait for the data
	\item Read data in the size of a $CHUNK$
	\item End of the reading process
	\item Computation of the data
	\item Write the processed SVs back to disk.
\end{dingautolist}

\begin{table}[htp]
\footnotesize
\vspace*{-0.6em}
\caption{35-qubit $H\_0$ time allocation (sec) working on SATA II and SATA III}\label{tab: gen_comp}
\vspace*{-0.5em}
\centering
    \begin{tabular}{|| c || c | c | c | c | c ||} 
        \hline
        & \ding{172} & \ding{173} & \ding{174} & \ding{175} & \ding{176}\\
        \hline
        SATA II & 14.6 & 74.87 & 0.57 & 89.19 & 67.83 \\
        \hline
        SATA III & 13.61 & 61.54 & 0.36 & 89.29 & 52.81 \\
        \hline
    \end{tabular}
\end{table}
In TABLE~\ref{tab: gen_comp}, we present the time allocation of $H\_0$ working on 35 qubits. The data transfer dominates the simulation, which takes $63.89\%$ and $58.97\%$, respectively, not to say the case of $*\_disk$ and $*\_system$ with two-times chunk access. Compared with SATA II, the SATA III accelerates the process of \ding{173} about $17.8\%$ and \ding{176} around $22.14\%$ in the case of $CHUNK = 4096$. On the other hand, we find out that the SATA generations do not affect the \ding{172} and, of course, \ding{175}, which is primarily related to the target FPGA.

\subsection{Qu-Trefoil Working on Large-scale Qubits Simulation}
Till now, the SV-based quantum circuit simulation over 35 qubits primarily depends on supercomputers, as explained in Section~\ref{sec: intro}. This section evaluates the Qu-Trefoil working on qubit scales over 40. To comprehensively assess the designed quantum gates, we still accept $H$ as our target in this part. The evaluation runs separately under SATA II and SATA III, setting $CHUNK=4096$. 

\begin{figure}[htp]
    \centering
    \includegraphics[width=0.9\linewidth]{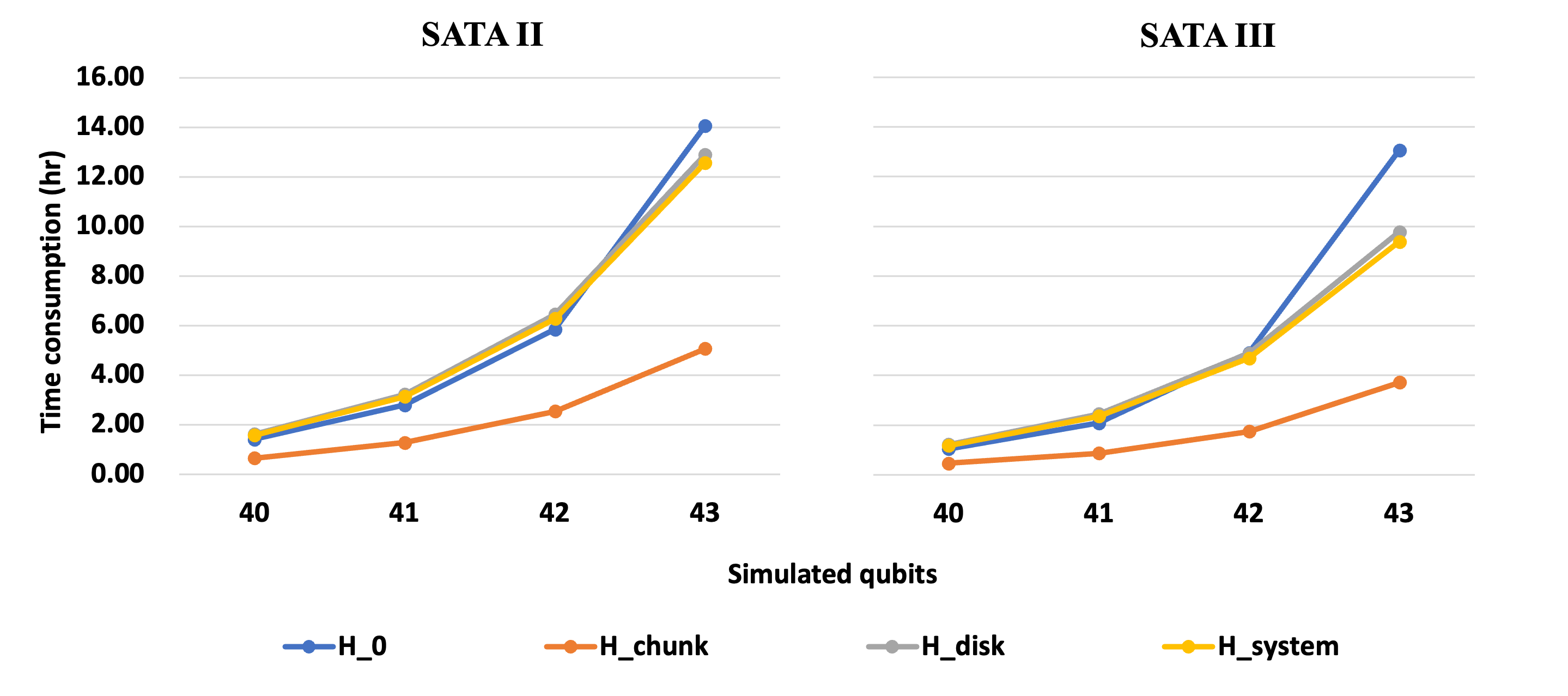}
    \caption{Exponential increase in time consumption of Qu-Trefoil on $H$ gate}
    \label{fig: large_scale}
\end{figure}

As shown in Fig.\ref{fig: large_scale}, regardless of whether SATA II or SATA III is used, the time consumption increases near-exponentially with the number of simulated qubits due to the more significant amounts of state vectors (SVs). The detailed time consumption data is also provided in TABLE\ref{tab: large_scale}. Consequently, depending on the position of the target qubit, the proposed simulator can perform a 43-qubit simulation for the $H$ gate in approximately 3.72 to 13.06 hours, with an average of 8.98 hours, which includes the most time-consuming case of quantum gate simulation referred to as $H\_system$.

\begin{table}[htp]
\footnotesize
\vspace*{-0.6em}
\caption{\textbf{Time consumption of $H$ gate simulation over 40 qubits on Qu-Trefoil}}\label{tab: large_scale}
\vspace*{-0.5em}
\centering
    \begin{tabular}{|| c || c | c | c | c ||}
    \hline
    \multicolumn{5}{||c||}{\textbf{Time consumption at SATA II (sec)}} \\
    \hline
    & 40 qubits & 41 qubits & 42 qubits & 43 qubits \\
    \hline
    $\bm{H\_0}$ & 5051.75 & 10048.91 & 21045.13 & 50630.67\\
    \hline
    $\bm{H\_chunk}$ & 2322.65 & 4587.63 & 9145.81 & 18223.82\\
    \hline
    $\bm{H\_disk}$ & 5801.09 & 11601.45 & 23202.64 & 46443.4\\
    \hline
    $\bm{H\_system}$ & 5662.45 & 11288.67 & 22591.53 & 45236.68\\
    \hline
    \multicolumn{5}{||c||}{\textbf{Time consumption at SATA III (sec)}} \\
    \hline
    & 40 qubits & 41 qubits & 42 qubits & 43 qubits \\
    \hline
    $\bm{H\_0}$ & 3758.50 & 7522.95 & 17673.06 & 47017.42\\
    \hline
    $\bm{H\_chunk}$ & 1699.98 & 3157.74 & 6279.13 & 13379.05\\
    \hline
    $\bm{H\_disk}$ & 4421.22 & 8813.50 & 17606.29 & 35198.60\\
    \hline
    $\bm{H\_system}$ & 4249.35 & 8485.18 & 16911.20 & 33786.68\\
    \hline
    \end{tabular}
\end{table}

\subsection{Comparison With State-of-the-art SV-based QCSs}
\begin{table*}[htp]
\footnotesize
\vspace*{-0.6em}
\caption{Comparison with state-of-the-art SV-based QCSs}\label{tab: compare}
\vspace*{-0.5em}
\centering
    \begin{tabular}{|| c || c | c | c | c | c | c | c ||}
    \hline
    \multirow{2}{*}{} & \multirow{2}{*}{\textbf{Simulator}} & \multirow{2}{*}{\textbf{Algorithm}} & \multirow{2}{*}{\textbf{Platform}} & \multicolumn{2}{c|}{\textbf{Performance}} & \multirow{2}{*}{\textbf{Cost (M\$)}} & \multirow{2}{*}{\textbf{Energy (MJ)}}\\
    \cline{5-6}
    & & & & \textbf{Qubits} & \textbf{Time (sec)} & &\\
    \hline
    \cite{fugaku} & Qulacs & Hadamard gate & Fugaku &  48 & 172.04 & 1,200 & 4,516\\
    \hline
    \cite{archer} & QuEST & Quantum Fourier Transform & ARCHER2 &  44 & 285 & 102 & 431\\
    \hline
    \cite{cuQuantum} & Qiskit Aer & Quantum Phase Estimation & Selene & 40 & 35 & 85 & 1,330\\
    \hline
    \cite{nvidia} & cuStateVec & Hadamard gate & NVIDIA DGX H100 & 33 & 3.57 & 0.27 & 0.036 \\
    \hline
    ours & Qu-Trefoil & Hadamard gate & Trefoil storage subsystem & 43 & 32345.44 & 0.042 & 0.4\\
    \hline
    ours (estimated) & Qu-Trefoil & Hadamard gate & Trefoil full system & 46 & $< 40000$ & $< 0.4$ & $< 4$\\
    \hline
    \end{tabular}
\end{table*}

\begin{table}[htp]
\footnotesize
\vspace*{-0.6em}
\caption{\textbf{Cost Breakdown for the Trefoil Platform}}\label{tab: cost_breakdown}
\vspace*{-0.5em}
\centering
    \begin{tabular}{|| c || c | c | c ||}
    \hline
    \multicolumn{4}{||c||}{\textbf{Trefoil Storage Subsystem}} \\
    \hline
    & Amount & Unit Cost (\$)& Total Cost (\$)\\
    \hline
    Subsystem FPGA & 1 & 12,616 & 12,616 \\
    \hline
    8TB SATA Disk & 32 & 689 & 22,048 \\
    \hline
    SATA Connector & 1 & 7,750 & 7,750 \\
    \hline
    \multicolumn{4}{||c||}{\textbf{Trefoil Full System}} \\
    \hline
    & Amount & Unit Cost (\$)& Total Cost (\$)\\
    \hline
    Storage Subsystem & 8 & 42,414 & 339,312 \\
    \hline
    Main System FPGA & 1 & 21,612 & 21,612 \\
    \hline
    \end{tabular}
\end{table}
In TABLE~\ref{tab: compare}, we present state-of-the-art implementations of SV-based QCSs, highlighting the simulators used, platforms, best-effort qubit scales, performance, costs, and energy consumption. For a GPU server like the NVIDIA H100, the qubit scale is still limited to 33 qubits due to memory constraints despite the fast simulation speed of just 3.57 seconds for an H gate. In contrast, only supercomputers such as Fugaku, ARCHER2, and NVIDIA Selene can simulate qubit scales exceeding 35 qubits on much more complex quantum circuits, such as the Quantum Fourier Transform and Quantum Phase Estimation. However, these systems come with high financial costs and energy consumption, requiring using the entire supercomputer as a simulator, making them infeasible for most researchers.

In our design, while Qu-Trefoil cannot match the processing speed of supercomputers or GPU servers, it offers an SV-based QCS platform for large-scale qubit simulation that is more accessible, flexible, and cost-effective than other simulators in initial investment and long-term maintenance. Specifically, the detailed cost breakdown in TABLE ~\ref{tab: cost_breakdown} provides a transparent view of the system's construction and highlights the flexibility of Trefoil. As a multi-FPGA storage system, Trefoil allows for modular component combinations, enabling users to configure the system based on their budget constraints. It is an appealing option for researchers and institutions with limited access to HPC resources, such as supercomputers and GPU clusters.

In terms of controlling the system, once the corresponding bitstream for a quantum gate, generated using Vivado 2022.2.2 based on the quantum circuit design, is loaded, necessary parameters are input via a Raspberry Pi 3 to control the target system. However, Qu-Trefoil requires reconfiguring the platform for each gate in the circuit, with the resulting state vectors saved to SATA disks after each reconfiguration. As outlined in Section~\ref{sec: intro}, Trefoil's main system can connect directly to eight storage subsystems, allowing it to handle simulations of up to 46 qubits without compromising performance speed at the current scale.

%% file: Contents/6_con.tex
\section{Conclusion}
\label{sec: con}

In this paper, we have proposed the world's first large-scale SV-based QCS, Qu-Trefoil, working on FPGA, targeting the jobs that can only finished by supercomputers so far. By delicately designing the quantum gates IPs using HLS, including $H$, $Z$, $S$, $CNOT$, $T$, and unitary matrix, the proposed system can construct any quantum circuit with quantum gates combination. Because of the limited on-chip resources, we carefully allocate the resources, taking the balance of resource utilization and performance. Due to the restricted speed of SATA disks, we utilize the pipeline in data processing to boost the system's performance.

In evaluating Qu-Trefoil, we comprehensively assess the proposed system's performance, including the designed quantum gate IPs' performance, chunk size extension in burst mode, SATA disk expansion, and the SATA generation influence on Qu-Trefoil. Finally, we conducted simulations on a qubit scale exceeding 40 using the H gate on a Trefoil storage subsystem. While we observed an exponential rise in time consumption corresponding to the qubit size, we can further expand the qubit scale by integrating with the main system without compromising speed, as explained. 

From the perspective of further system optimization, we can enhance performance through three approaches. First, regarding bandwidth, we can improve system efficiency by employing data compression techniques on state vectors (SVs) to increase the adequate bandwidth of the SATA disks, as demonstrated in our initial trial \cite{fpga_compressor}. Additionally, we will explore the impact of storing state vectors in 32-bit, 16-bit, and 16-bit bfloat (brain floating-point) formats instead of the 64-bit double-precision floating-point format on the QCS. Furthermore, in terms of the hardware platform, according to PHISON's announcement \cite{nvme_disk}, newly released NVMe SSDs offer over 100TB of storage with significantly faster data access speeds, which suggests that upgrading from SATA to an NVMe interface could enable Qu-Trefoil to support larger-scale quantum circuit simulations with vastly improved data communication efficiency. Finally, from an architectural standpoint, computation-in-memory (CIM) using flash memory \cite{CIM} is an emerging technology that may help address limitations of unfavorable data access patterns by embedding the circuit of generating addresses and creating chunks. Furthermore, CIM also allows for the parallelism of quantum gate operations across different memory cells, further improving the performance of Qu-Trefoil.

%% file: Contents/ack.tex
\section*{Acknowledgment}
This work was supported by the following sources: JST Grant Number JPMJPF2221 / JPMJCR19K1, JSPS KAKENHI Grant Number JP21H04869 / JP23K19979, KIOXIA Corporation, and Tsukuba Innovation Arena (TIA).

%% file: Contents/bio.tex
%

\begin{IEEEbiography}[
{\includegraphics[width=1in,height=1.25in,clip,keepaspectratio]{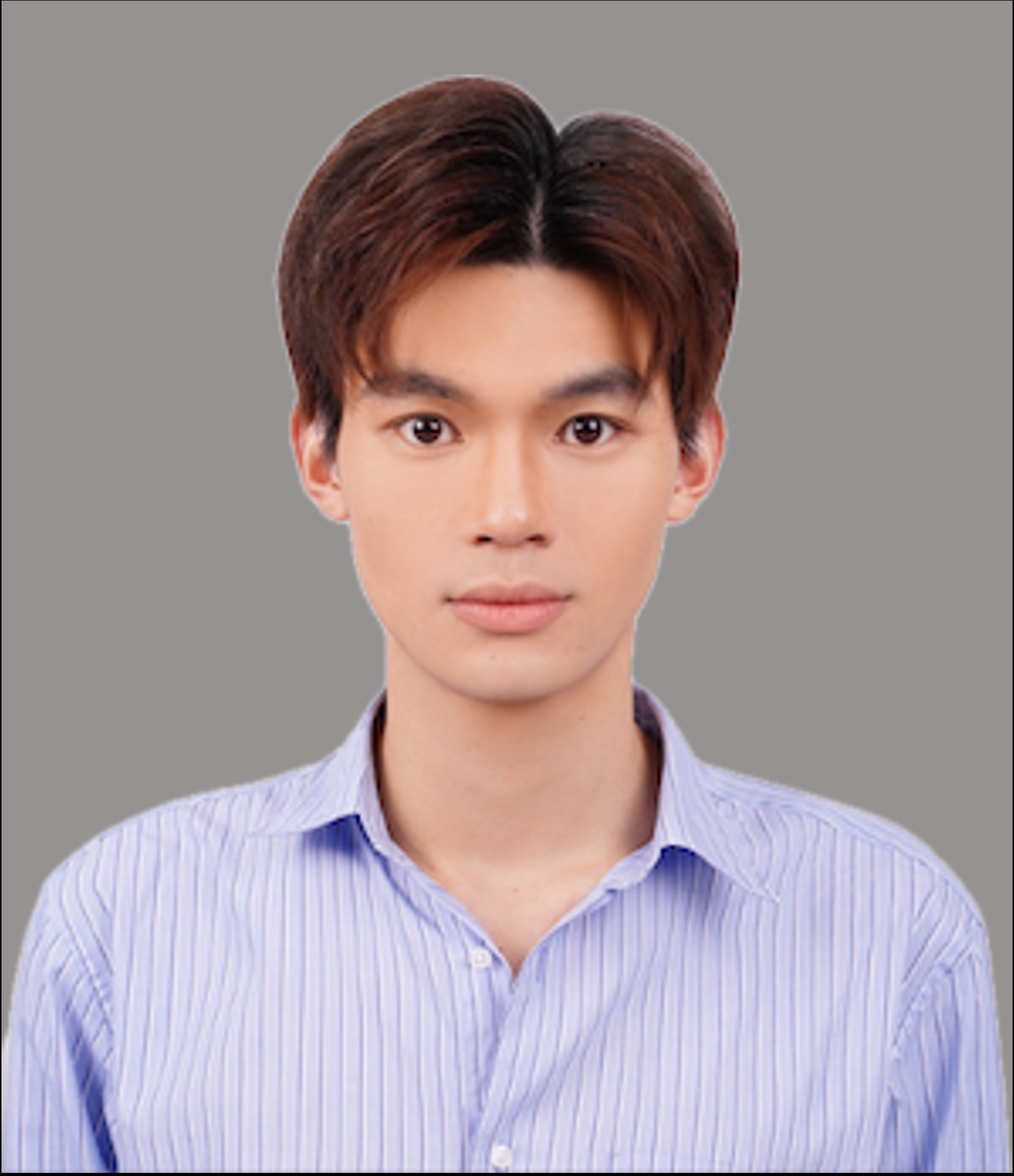}}
]{Kaijie Wei} (Member, IEEE) received the M.E. and Ph.D. degrees in Engineering from Keio University, Japan, in 2020 and 2023, respectively. He is currently a Project Assistant Professor at the Center for Sustainable Quantum AI (SQAI), Keio University. His research interests primarily concern computer architecture and system optimization, particularly reconfigurable devices.
\end{IEEEbiography}
\vspace{-5mm}

\begin{IEEEbiography}[
{\includegraphics[width=1in,height=1.25in,clip,keepaspectratio]{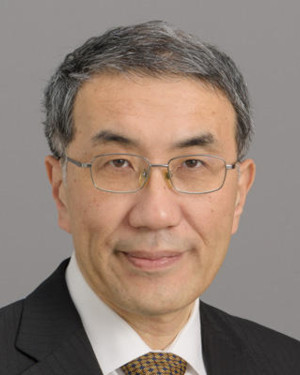}}
]{Hideharu Amano} (Life Member, IEEE) received a Ph.D. degree in electrical engineering from Keio University, Tokyo, Japan, in 1986. He is currently a Senior Fellow with the Systems Design Lab (d.lab), The University of Tokyo. His research interests include parallel architecture and reconfigurable computing.
\end{IEEEbiography}
\vspace{-5mm}

\begin{IEEEbiography}[
{\includegraphics[width=1in,height=1.25in,clip,keepaspectratio]{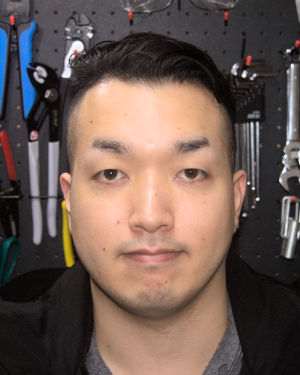}}
]{Ryohei Niwase} (Graduate Student Member, IEEE) received an MEng degree from the Graduate School of Science and Technology, Tsukuba University, Japan, in 2022 and is currently pursuing PhD at the same university. He has worked in signal processing and instrumentation engineering, and his current research interests cover the areas of numerical computing with FPGA clusters.
\end{IEEEbiography}
\vspace{-5mm}

\begin{IEEEbiography}
[{\includegraphics[width=1in,height=1.25in,clip,keepaspectratio]{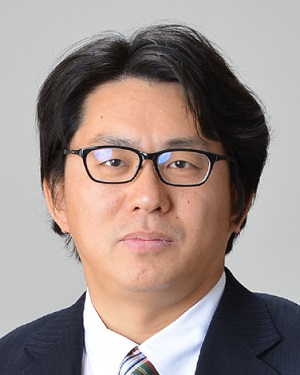}}]
{Yoshiki Yamgauchi} (Member, IEEE) received his PhD from the University of Tsukuba in 2003. He joined the University of Tsukuba in 2005 as an assistant professor. He progressed to an associate professor and a professor in 2015 and 2024, respectively. Since July 2011, he has contributed as a collaborative fellow at the Center for Computational Science at the University of Tsukuba. He has also been a professor at the Research and Education Center for Semiconductor and Digital Technologies at Kumamoto University since June 2023.
\end{IEEEbiography}
\vspace{-5mm}

\begin{IEEEbiography}
[{\includegraphics[width=1in,height=1.25in,clip,keepaspectratio]{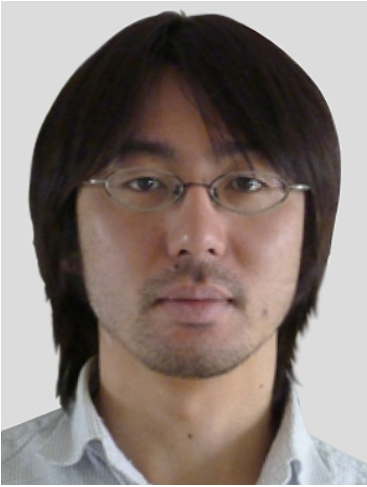}}]
{Takefumi Miyoshi} received his B.E., M.E., and D.E. from Tokyo Institute of Technology in 2003, 2005, and 2007, respectively. Since April 2010, he has been an assistant professor at the Graduate School of Information Systems, UEC. His research interests are compiler techniques, many-core processor architecture, and hardware and software co-design. He is a member of the ACM, IEICE, and IPSJ.
\end{IEEEbiography}




%% file: main.bbl
\begin{thebibliography}{10}
\providecommand{\url}[1]{#1}
\csname url@samestyle\endcsname
\providecommand{\newblock}{\relax}
\providecommand{\bibinfo}[2]{#2}
\providecommand{\BIBentrySTDinterwordspacing}{\spaceskip=0pt\relax}
\providecommand{\BIBentryALTinterwordstretchfactor}{4}
\providecommand{\BIBentryALTinterwordspacing}{\spaceskip=\fontdimen2\font plus
\BIBentryALTinterwordstretchfactor\fontdimen3\font minus
  \fontdimen4\font\relax}
\providecommand{\BIBforeignlanguage}[2]{{%
\expandafter\ifx\csname l@#1\endcsname\relax
\typeout{** WARNING: IEEEtran.bst: No hyphenation pattern has been}%
\typeout{** loaded for the language `#1'. Using the pattern for}%
\typeout{** the default language instead.}%
\else
\language=\csname l@#1\endcsname
\fi
#2}}
\providecommand{\BIBdecl}{\relax}
\BIBdecl

\bibitem{shor}
P.~Shor, ``Algorithms for quantum computation: discrete logarithms and
  factoring,'' in \emph{Proceedings 35th Annual Symposium on Foundations of
  Computer Science}, 1994, pp. 124--134.

\bibitem{grover}
L.~K. Grover, ``A fast quantum mechanical algorithm for database search,'' in
  \emph{Proceedings of the twenty-eighth annual ACM symposium on Theory of
  computing}, 1996, pp. 212--219.

\bibitem{QPE}
\BIBentryALTinterwordspacing
H.~Ni, H.~Li, and L.~Ying, ``On low-depth algorithms for quantum phase
  estimation,'' \emph{Quantum}, vol.~7, p. 1165, Nov. 2023. [Online].
  Available: \url{http://dx.doi.org/10.22331/q-2023-11-06-1165}
\BIBentrySTDinterwordspacing

\bibitem{ibm_latest}
M.~Swayne, ``Ibm reportedly partnering with japan’s aist to develop
  10,000-qubit quantum computer,''
  \url{https://thequantuminsider.com/2024/06/18/ibm-reportedly-partnering-with-japans-aist-to-develop-10000-qubit-quantum-computer/},
  2024, [Accessed 12-09-2024].

\bibitem{quantum_challenegs}
\BIBentryALTinterwordspacing
S.~S. Gill, O.~Cetinkaya, S.~Marrone, D.~Claudino, D.~Haunschild, L.~Schlote,
  H.~Wu, C.~Ottaviani, X.~Liu, S.~P. Machupalli, K.~Kaur, P.~Arora, J.~Liu,
  A.~Farouk, H.~H. Song, S.~Uhlig, and K.~Ramamohanarao, ``Quantum computing:
  Vision and challenges,'' 2024. [Online]. Available:
  \url{https://arxiv.org/abs/2403.02240}
\BIBentrySTDinterwordspacing

\bibitem{DM_intro}
\BIBentryALTinterwordspacing
A.~D. Patel, ``The quantum density matrix and its many uses: From quantum
  structure to quantum chaos and noisy simulators,'' \emph{Journal of the
  Indian Institute of Science}, vol. 103, no.~2, p. 401–417, Apr. 2023.
  [Online]. Available: \url{http://dx.doi.org/10.1007/s41745-023-00406-4}
\BIBentrySTDinterwordspacing

\bibitem{TN_intro}
\BIBentryALTinterwordspacing
T.~B. Wahl and S.~Strelchuk, ``Simulating quantum circuits using efficient
  tensor network contraction algorithms with subexponential upper bound,''
  \emph{Phys. Rev. Lett.}, vol. 131, p. 180601, Oct 2023. [Online]. Available:
  \url{https://link.aps.org/doi/10.1103/PhysRevLett.131.180601}
\BIBentrySTDinterwordspacing

\bibitem{SV_app_1}
\BIBentryALTinterwordspacing
B.~Fang, M.~Y. Ozkaya, A.~Li, U.~V. Catalyurek, and S.~Krishnamoorthy, ``{
  Efficient Hierarchical State Vector Simulation of Quantum Circuits via
  Acyclic Graph Partitioning },'' in \emph{2022 IEEE International Conference
  on Cluster Computing (CLUSTER)}.\hskip 1em plus 0.5em minus 0.4em\relax Los
  Alamitos, CA, USA: IEEE Computer Society, Sep. 2022, pp. 289--300. [Online].
  Available:
  \url{https://doi.ieeecomputersociety.org/10.1109/CLUSTER51413.2022.00041}
\BIBentrySTDinterwordspacing

\bibitem{fugaku}
\BIBentryALTinterwordspacing
H.~De~Raedt, F.~Jin, D.~Willsch, M.~Willsch, N.~Yoshioka, N.~Ito, S.~Yuan, and
  K.~Michielsen, ``Massively parallel quantum computer simulator, eleven years
  later,'' \emph{Computer Physics Communications}, vol. 237, p. 47–61, Apr.
  2019. [Online]. Available: \url{http://dx.doi.org/10.1016/j.cpc.2018.11.005}
\BIBentrySTDinterwordspacing

\bibitem{archer}
\BIBentryALTinterwordspacing
J.~Adamski, J.~P. Richings, and O.~T. Brown, ``Energy efficiency of quantum
  statevector simulation at scale,'' in \emph{Proceedings of the SC ’23
  Workshops of The International Conference on High Performance Computing,
  Network, Storage, and Analysis}, ser. SC-W 2023.\hskip 1em plus 0.5em minus
  0.4em\relax ACM, Nov. 2023. [Online]. Available:
  \url{http://dx.doi.org/10.1145/3624062.3624270}
\BIBentrySTDinterwordspacing

\bibitem{cuQuantum}
\BIBentryALTinterwordspacing
H.~Bayraktar, A.~Charara, D.~Clark, S.~Cohen, T.~Costa, Y.~L. Fang, Y.~Gao,
  J.~Guan, J.~Gunnels, A.~Haidar, A.~Hehn, M.~Hohnerbach, M.~Jones, T.~Lubowe,
  D.~Lyakh, S.~Morino, P.~Springer, S.~Stanwyck, I.~Terentyev, S.~Varadhan,
  J.~Wong, and T.~Yamaguchi, ``cuquantum sdk: A high-performance library for
  accelerating quantum science,'' in \emph{2023 IEEE International Conference
  on Quantum Computing and Engineering (QCE)}.\hskip 1em plus 0.5em minus
  0.4em\relax Los Alamitos, CA, USA: IEEE Computer Society, sep 2023, pp.
  1050--1061. [Online]. Available:
  \url{https://doi.ieeecomputersociety.org/10.1109/QCE57702.2023.00119}
\BIBentrySTDinterwordspacing

\bibitem{nvidia}
``{cuQuantum Accelerate quantum computing research.}, howpublished =
  {\url{https://developer.nvidia.com/cuquantum-sdk}}, note = {Accessed:
  2024-02-05}.''

\bibitem{HBM_1}
Z.~Wang, H.~Huang, J.~Zhang, and G.~Alonso, ``Shuhai: Benchmarking high
  bandwidth memory on fpgas,'' in \emph{2020 IEEE 28th Annual International
  Symposium on Field-Programmable Custom Computing Machines (FCCM)}, 2020, pp.
  111--119.

\bibitem{HBM_2}
H.~Huang, Z.~Wang, J.~Zhang, Z.~He, C.~Wu, J.~Xiao, and G.~Alonso, ``Shuhai: A
  tool for benchmarking high bandwidth memory on fpgas,'' \emph{IEEE
  Transactions on Computers}, vol.~71, no.~5, pp. 1133--1144, 2022.

\bibitem{xilinx}
``{Ultrascale plus fpga product selection guide}, howpublished =
  {\url{https://docs.xilinx.com/v/u/en-us/ultrascale-plus-fpga-product-selection-guide/}},
  note = {Accessed: 2024-02-08}.''

\bibitem{IntelAgilex}
M.~van~der Zalm, ``{I}ntel {A}gilex® 5 {F}{P}{G}{A}s and {S}o{C} {F}{P}{G}{A}s
  {A}re {I}deal for {M}idrange {A}pplications {R}equiring {H}igher
  {P}erformance, {L}ower {P}ower, and {S}maller {F}orm {F}actors ---
  intel.com,''
  \url{https://www.intel.com/content/www/us/en/content-details/764697/intel-agilex-5-fpgas-and-soc-fpgas-are-ideal-for-midrange-applications-requiring-higher-performance-lower-power-and-smaller-form-factors.html},
  26-06-2023, [Accessed 08-02-2024].

\bibitem{MCSOC}
R.~Niwase, H.~Harasawa, Y.~Yamaguchi, W.~Kaijie, and H.~Amano, ``A cost/power
  efficient storage system with directly connected fpga and sata disks,'' in
  \emph{2023 IEEE 16th International Symposium on Embedded Multicore/Many-core
  Systems-on-Chip (MCSoC)}, 2023, pp. 51--58.

\bibitem{Qulacs}
\BIBentryALTinterwordspacing
Y.~Suzuki, Y.~Kawase, Y.~Masumura, Y.~Hiraga, M.~Nakadai, J.~Chen, K.~M.
  Nakanishi, K.~Mitarai, R.~Imai, S.~Tamiya, T.~Yamamoto, T.~Yan, T.~Kawakubo,
  Y.~O. Nakagawa, Y.~Ibe, Y.~Zhang, H.~Yamashita, H.~Yoshimura, A.~Hayashi, and
  K.~Fujii, ``Qulacs: a fast and versatile quantum circuit simulator for
  research purpose,'' \emph{Quantum}, vol.~5, p. 559, Oct. 2021. [Online].
  Available: \url{http://dx.doi.org/10.22331/q-2021-10-06-559}
\BIBentrySTDinterwordspacing

\bibitem{Q_app}
\BIBentryALTinterwordspacing
F.~Bova, A.~Goldfarb, and R.~G. Melko, ``Commercial applications of quantum
  computing,'' \emph{EPJ Quantum Technology}, vol.~8, no.~1, p.~2, 2021.
  [Online]. Available: \url{https://doi.org/10.1140/epjqt/s40507-021-00091-1}
\BIBentrySTDinterwordspacing

\bibitem{DM_app_1}
\BIBentryALTinterwordspacing
A.~D. Patel, ``The quantum density matrix and its many uses: From quantum
  structure to quantum chaos and noisy simulators,'' \emph{Journal of the
  Indian Institute of Science}, vol. 103, no.~2, p. 401–417, Apr. 2023.
  [Online]. Available: \url{http://dx.doi.org/10.1007/s41745-023-00406-4}
\BIBentrySTDinterwordspacing

\bibitem{TN_app_1}
\BIBentryALTinterwordspacing
T.~Nguyen, D.~Lyakh, E.~Dumitrescu, D.~Clark, J.~Larkin, and A.~McCaskey,
  ``Tensor network quantum virtual machine for simulating quantum circuits at
  exascale,'' \emph{ACM Transactions on Quantum Computing}, vol.~4, no.~1, pp.
  1--21, Oct. 2022. [Online]. Available: \url{https://doi.org/10.1145/3547334}
\BIBentrySTDinterwordspacing

\bibitem{IQS}
\BIBentryALTinterwordspacing
G.~G. Guerreschi, J.~Hogaboam, F.~Baruffa, and N.~P.~D. Sawaya, ``Intel quantum
  simulator: a cloud-ready high-performance simulator of quantum circuits,''
  \emph{Quantum Science and Technology}, vol.~5, no.~3, p. 034007, May 2020.
  [Online]. Available: \url{http://dx.doi.org/10.1088/2058-9565/ab8505}
\BIBentrySTDinterwordspacing

\bibitem{quest}
\BIBentryALTinterwordspacing
T.~Jones, A.~Brown, I.~Bush, and S.~C. Benjamin, ``Quest and high performance
  simulation of quantum computers,'' \emph{Scientific Reports}, vol.~9, no.~1,
  pp. 1--11, 2019. [Online]. Available:
  \url{https://doi.org/10.1038/s41598-019-47174-9}
\BIBentrySTDinterwordspacing

\bibitem{annealing_1}
E.~Forno, A.~Acquaviva, Y.~Kobayashi, E.~Macii, and G.~Urgese, ``A parallel
  hardware architecture for quantum annealing algorithm acceleration,'' in
  \emph{2018 IFIP/IEEE International Conference on Very Large Scale Integration
  (VLSI-SoC)}, 2018, pp. 31--36.

\bibitem{annealing_2}
H.~M. Waidyasooriya and M.~Hariyama, ``{Highly-Parallel FPGA Accelerator for
  Simulated Quantum Annealing},'' \emph{IEEE Tran. on Emerging Topics in
  Computing}, vol.~9, no.~4, pp. 2019--2029, 2021.

\bibitem{Ising}
\BIBentryALTinterwordspacing
A.~Mondal and A.~Srivastava, ``{Ising-FPGA: A Spintronics-Based Reconfigurable
  Ising Model Solver},'' \emph{ACM Transactions on Design Automation of
  Electronic Systems}, vol.~26, no.~1, pp. 1--27, sep 2020. [Online].
  Available: \url{https://doi.org/10.1145/3411511}
\BIBentrySTDinterwordspacing

\bibitem{FPGA_Qcomp_1}
Y.~Jungjarassub and K.~Piromsopa, ``{A Performance Optimization of Quantum
  Computing Simulation using FPGA},'' in \emph{ECTI-CON 2022}, 2022, pp. 1--4.

\bibitem{FPGA_Qcomp_2}
Y.~Hong \emph{et~al.}, ``{Implementation of a Quantum Circuit Simulator using
  Classical Bits},'' in \emph{2022 IEEE AICAS}, 2022, pp. 472--474.

\bibitem{FPT_1}
R.~Niwase, H.~Harasawa, Y.~Yamaguchi, W.~Kaijie, and H.~Amano, ``A cost/power
  efficient storage system with directly connected fpga and sata disks,'' in
  \emph{2023 IEEE 16th International Symposium on Embedded Multicore/Many-core
  Systems-on-Chip (MCSoC)}, 2023, pp. 51--58.

\bibitem{FPT_2}
K.~Wei, R.~Niwase, H.~Amano, Y.~Yamaguchi, and T.~Miyoshi, ``A state vector
  quantum simulator working on fpgas with extensible sata storage,'' in
  \emph{2023 International Conference on Field Programmable Technology
  (ICFPT)}, 2023, pp. 272--273.

\bibitem{mpiQulacs}
A.~Tabuchi, S.~Imamura, M.~Yamazaki, T.~Honda, A.~Kasagi, H.~Nakao,
  N.~Fukumoto, and K.~Nakashima, ``mpiqulacs: A scalable distributed quantum
  computer simulator for arm-based clusters,'' in \emph{2023 IEEE International
  Conference on Quantum Computing and Engineering (QCE)}, vol.~01, 2023, pp.
  959--969.

\bibitem{trefoil}
N.~Umezu, Y.~Yamaguchi, and T.~Boku, ``An fpga-based storage control with load
  balancing,'' in \emph{2021 IEEE International Conference on Cluster Computing
  (CLUSTER)}, 2021, pp. 791--794.

\bibitem{disk_comparsion}
Y.~Son, H.~Kang, H.~Han, and H.~Y. Yeom, ``An empirical evaluation of nvm
  express ssd,'' in \emph{2015 International Conference on Cloud and Autonomic
  Computing}, 2015, pp. 275--282.

\bibitem{samsung_ssd}
{Samsung Electronics Co., Ltd.}, ``{870 QVO Data Sheet Version 1.1},''
  \url{https://download.semiconductor.samsung.com/resources/data-sheet/Samsung\_SSD\_870\_QVO\_Data\_Sheet\_Rev1.1.pdf},
  2020, accessed: 2024-09-10.

\bibitem{litex}
\BIBentryALTinterwordspacing
F.~Kermarrec, S.~Bourdeauducq, J.~L. Lann, and H.~Badier, ``Litex: an
  open-source soc builder and library based on migen python {DSL},''
  \emph{CoRR}, vol. abs/2005.02506, 2020. [Online]. Available:
  \url{https://arxiv.org/abs/2005.02506}
\BIBentrySTDinterwordspacing

\bibitem{litex_github}
F.~Kermarrec, ``Litex: An open-source soc builder and library based on migen
  python dsl,'' \url{https://github.com/enjoy-digital/litex}, 2020, accessed:
  2024-09-06.

\bibitem{migen_github}
S.~Bourdeauducq, ``Migen: A python toolbox for building complex digital
  hardware,'' \url{https://github.com/m-labs/migen}, 2019, accessed:
  2024-09-06.

\bibitem{fpga_compressor}
\BIBentryALTinterwordspacing
K.~Wei, H.~Amano, R.~Niwase, and Y.~Yamaguchi, ``A data compressor for
  fpga-based state vector quantum simulators,'' in \emph{Proceedings of the
  14th International Symposium on Highly Efficient Accelerators and
  Reconfigurable Technologies, {HEART} 2024, Porto, Portugal, June 19-21,
  2024}, L.~Josipovic, P.~Zhou, S.~Shanker, J.~M.~P. Cardoso, J.~Anderson, and
  S.~Yuichiro, Eds.\hskip 1em plus 0.5em minus 0.4em\relax {ACM}, 2024, pp.
  63--70. [Online]. Available: \url{https://doi.org/10.1145/3665283.3665293}
\BIBentrySTDinterwordspacing

\bibitem{nvme_disk}
C.~Robinson, ``Phison pascari d200v pcie gen5 nvme with 122.88tb of capacity
  announced,''
  \url{https://www.servethehome.com/phison-pascari-d200v-pcie-gen5-nvme-ssd-with-122-88tb-of-capacity-announced/},
  2024, [Accessed 12-09-2024].

\bibitem{CIM}
\BIBentryALTinterwordspacing
Y.~Halawani and B.~Mohammad, \emph{In-Memory Computing Using FLASH
  Memory}.\hskip 1em plus 0.5em minus 0.4em\relax Cham: Springer Nature
  Switzerland, 2024, pp. 123--125. [Online]. Available:
  \url{https://doi.org/10.1007/978-3-031-34233-2\_7}
\BIBentrySTDinterwordspacing

\end{thebibliography}
